\documentclass[aps,prb,twocolumn,floatfix,superscriptaddress,amsmath,amssymb]{revtex4}

\usepackage[final]{graphicx}
\usepackage{bbold,bm}

\newcommand\vex[1]{\mathbf{#1}}

\def\d{\mathrm{d}}
\def\id{\mathbb{1}}

\def\diag{\mathrm{diag}}
\def\ii{\mathrm{i}}
\def\RR{\mathrm{Re}}
\def\II{\mathrm{Im}}
\def\osum{{\circ\mkern-15mu \sum}}
\def\sslash#1{\setbox0=\hbox{$#1$}			
   \dimen0=\wd0                                		
   \setbox1=\hbox{/} \dimen1=\wd1  	 		
   \ifdim\dimen0>\dimen1                               	
      \rlap{\hbox to \dimen0{\hfil / \hfil}} 	  	
      #1                                      			
   \else                                        			
      \rlap{\hbox to \dimen1{\hfil$#1$\hfil}}   		
      \hbox{/} 	                              			
   \fi}   
\def\slash#1{\hbox{$#1$\kern-0.35em\raise0.1ex\hbo+
{/}}}

\begin{document}

\title{Creation and manipulation of anyons in a layered superconductor-2DEG system}

\author{G. Rosenberg}
\affiliation{Department of Physics and Astronomy,
University of British Columbia, Vancouver, BC, Canada V6T 1Z1}

\author{B. Seradjeh}
\affiliation{Department of Physics and Astronomy,
University of British Columbia, Vancouver, BC, Canada V6T 1Z1}
\affiliation{Department of Physics,
University of Illinois, 1110 West Green St, Urbana, IL 61801-3080}

\author{C. Weeks}
\affiliation{Department of Physics and Astronomy,
University of British Columbia, Vancouver, BC, Canada V6T 1Z1}

\author{M. Franz}
\affiliation{Department of Physics and Astronomy,
University of British Columbia, Vancouver, BC, Canada V6T 1Z1}

\date{\today}

\begin{abstract}
We describe and analyze in detail our recent theoretical proposal for the realization and manipulation of anyons in a weakly interacting system consisting of a two-dimensional electron gas in the {\em integer} quantum Hall regime adjacent to a type-II superconducting film with an artificial array of pinning sites. The anyon is realized in response to a defect in the pinned vortex lattice and carries a charge $\pm e/2$ and a statistical angle $\pi/4$. We establish this result, both analytically and numerically, in three complementary approaches: (i) a continuum model of two-dimensional electrons in the vortex lattice of the superconducting film; (ii) a minimal tight-binding lattice model that captures the essential features of the system; and (iii) an effective theory of the superconducting vortex lattice superposed on the integer quantum Hall state. We propose a novel method to measure the fractional charge \emph{directly} in a bulk transport experiment and an \emph{all-electric} setup for an ``anyon shuttle'' implementing the braiding operations. We briefly discuss conditions for fabricating the system in the lab and its potential applications in quantum information processing with non-Abelian anyons.
\end{abstract}
\maketitle

\section{Introduction}

In two dimensions (2D) the wavefunction of indistinguishable particles can have exotic exchange properties: upon the exchange of two particles known as ``anyons''~\cite{Wil82a} it acquires a phase factor $\exp(\ii\theta)\neq\pm 1$ with an arbitrary statistical angle $\theta$. This is in contrast to the situation in higher dimensions where $\theta=0$ or $\pi$ corresponding to bosons or fermions, respectively. The reason for this departure from the usual quantum statistics arises from the unique topological properties of 2D systems.~\cite{LeiMyr77a} 
If the ground state of such identical particles is degenerate the exchange operations can be described by matrices in the degenerate subspace, which are generically non-Abelian. Such particles are referred to as non-Abelian anyons. It is also known that particles with fractional charge may arise, regardless of dimensionality, as the collective response of a many-body system to topologically nontrivial background fields.~\cite{JacReb76a,SuSchHee79a,GolWil81a}

The interest in anyons is not merely academic. The topological character of anyons means that local perturbations alter their properties exponentially weakly with a characteristic length of the order of the separation between anyons. (For a single anyon this length scale is replaced by the system size.) This ``topological protection'' has been argued to be useful for fault-tolerant quantum information processing.~\cite{Kit03a} If the anyons have a rich enough non-Abelian structure, one may realize a universal quantum computer, in which computations are performed by braiding anyons.~\cite{freedman1} Otherwise, topological protection for a non-universal subset of operations may be obtained. In any case, the system can be considered, at least, as a topologically protected quantum memory. However, the insensitivity to local perturbations also poses a problem for manipulating anyons, which is necessary for braiding operations, and for the storage and retrieval of information from the quantum memory. 

\begin{figure}[b]
\begin{center}
\includegraphics[width=3.45in]{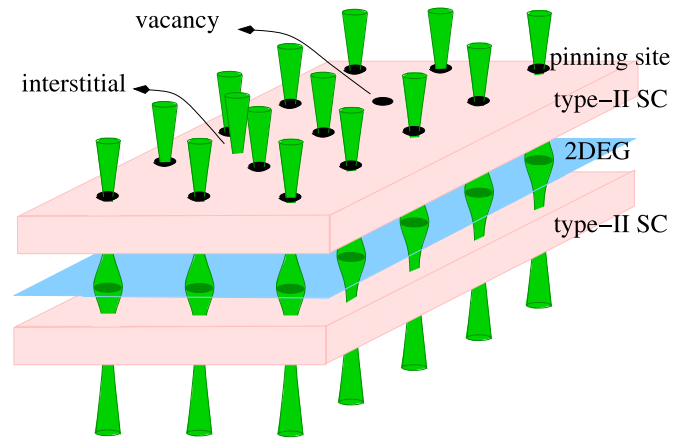}
\caption{(color online) The schematic diagram of the proposed device. A 2D electron gas is sandwiched between two type-II superconducting films.  }
\label{fig:sys}
\end{center}
\end{figure}

How could anyons be realized in nature? Of course, the first condition is the restriction to two spatial dimensions. The fact that the world is three-dimensional immediately implies that anyons can only be realized as collective excitations of a many-body system whose constituent particles are necessarily fermions or bosons. The emergence of such excitations constitutes the phenomenon of fractionalization in condensed-matter systems.

Wilczek~\cite{Wil82b} proposed a simple model of an anyon as a bound state of a charge $q$ and magnetic flux $\Phi$, in which either the charge or the flux, or both, have a fractional value in units of the electron charge, $e$, and flux quantum $\Phi_0=hc/e$, respectively. 
A closely related state is realized in a 2D electron gas (2DEG) in a perpendicular magnetic field in the fractional quantum Hall (FQH) regime, described by the filling factor $\nu$. When $1/\nu$ is an odd integer the excitations carry a fractional charge $\nu e$ (Ref.~\onlinecite{Lau83b}) and have a statistical angle $\nu\pi$ (Ref.~\onlinecite{AroSchWil84a}). 
FQH physics is the canonical example of fractionlaization.
The Coulomb interaction is understood to play an important role in stabilizing the FQH states, which may be  considered strongly correlated in the sense that they cannot be described by filling a set of single-particle states.

One may naturally ask, then, if this is a necessary condition of fractionalization. It has been known for a long time that strong correlations are not necessary for fractionalization of charge in one and three dimensions,
\cite{JacReb76a,SuSchHee79a,GolWil81a} where exchange statistics are trivial. In two dimensions, fractional charge and statistics have recently been argued to arise in certain lattice models~\cite{HouChaMud07a,SerWeeFra08a,ChaHouJac07a, SerFra08a} that preserve time-reversal symmetry and can be considered weakly-interacting. Furthermore, we proposed~\cite{WeeRosSer07a} a weakly-interacting system to realize anyons that can be described by a Slater determinant of single-particle states. This proposal has the potential to be useful for manipulating anyons. Unlike some other proposals it also has a realistic chance to be fabricated in the lab.

The system consists of a 2DEG in the \emph{integer} quantum Hall state adjacent to a thin slab of a type-II superconductor. The idea is to employ the quantization of flux by the superconductor in units of $\frac12\Phi_0$, i.e. \emph{half} the natural quantum of flux of the 2DEG, as well as the quantization of Hall conductance in the quantum Hall state of the 2DEG. A sketch of a possible arrangement is shown in Fig.~\ref{fig:sys}. We have chosen a symmetric placement of superconducting films around the 2DEG  in order to minimize the spread and the in-plane component of the vortex magnetic field at the 2DEG. This is not necessary if similar conditions can be obtained in a different geometry. Both of these systems admit an effective single-particle description. The integer quantum Hall state can be understood as a full Landau level of essentially free electrons. The superconductor may be described as a condensate of Cooper pairs. It is then interesting that the proximity of these two weakly-interacting systems results in the fractionalization of charge and statistics. Anyons are formed in response to defects in a pinned Abrikosov vortex lattice in the superconducting film. Such defects may be detected and manipulated by scanning superconducting quantum interference devices (SQUID) and Hall bar probes or magnetic tips in magnetic force microscopy that have been developed over the years.~\cite{KirTsuRup96a,AusLuaStr08a,BluSebGui06a,GarWynBon02a,StrHofAus08a} Therefore, our anyons have the additional potential of being manipulated despite their topological protection. As we shall describe in the conclusion a similar setup with the FQH state at $\nu=5/2$ should realize, and allow for the manipulation of, non-Abelian anyons. These anyons are robust as long as the spread of the vortex-lattice defect, controlled by the superconducting penetration depth and the distance between the two layers, is small relative to the distance between the defects.

In the present paper, in addition to providing the details and some extensions of our previous analysis,~\cite{WeeRosSer07a} we present new analytical and numerical work from alternative starting points that shed light on different aspects of the system. Moreover, we present a novel and concrete method to measure the fractional charge of anyons in a bulk transport experiment that utilizes the vortices in the superconducting film. We also describe an idea for the ``anyon shuttle'', a system to manipulate anyons bound to the flux defects by purely electrical means.

The rest of the paper is organized as follows. In Sec.~\ref{sec:sys} we describe the system and provide a general argument for our findings. In Sec.~\ref{sec:cont} we discuss a continuum approach to the electrodynamic response of the 2DEG to the vortex lattice and its defects, employing various simplifying assumptions and a combination of analytical and numerical methods. In Sec.~\ref{sec:latt} we formulate and study a minimal lattice model of the system by exact diagonalization numerically as well as in the continuum limit analytically. In Sec.~\ref{sec:eff} we study an effective theory of the system, which is expressed as a Chern-Simons--Maxwell theory that we use to study the interplay of the superconducting and the topological orders in the system. Given the interesting physics predicted here and the potential application of the system in manipulating anyons, we believe it is important to make an effort to fabricate it in the lab. Sec.~\ref{sec:exp} contains our proposal to measure the fractional charge and a description of the anyon shuttle. Finally, we conclude in Sec.~\ref{sec:conc} by discussing the conditions for the experimental realization of the system. Details of some calculations as well as the derivation of some known results are given in Appendices~\ref{app:Schr}--\ref{app:tknn} in order to make the paper self-contained.

\section{The system}\label{sec:sys}

The proposed system has two components: a layer hosting a 2DEG in an integer quantum Hall state, and a film of type-II superconductor. The purpose of the superconducting film is to quantize the magnetic flux into an Abrikosov vortex lattice, where each vortex carries a flux $\frac12\Phi_0$, which  plays a central role in the physics we describe. Cooper pair condensation in the superconducting film ensures that this quantization is extremely precise. 

As mentioned above, the anyon is to be realized in the 2DEG in response to a defect, i.e. an extra or missing $\frac12\Phi_0$ flux in the vortex lattice. However, a naturally formed Abrikosov lattice is (almost) incompressible.~\cite{ConSch74a} This means that if defects are to be created by moving one of the vortices, the vortex lattice will rearrange itself in such a way as to compensate for the additional or missing flux in the corresponding regions. Thus, in order to allow for the creation of such defects, we propose to artificially imprint an array of pinning sites on the superconducting film. The pinning sites are regions where superconducting order is weakened and can be created in a variety of ways.~\cite{WelXiaNov05a} They attract and pin superconducting vortices, thereby preventing the incompressible rearrangement of vortices in response to a defect. At the matching field $B_M$ the number of vortices equals the number of pinning sites. An increase (decrease) in the field away from $B_M$ will induce a corresponding number of interstitial (vacancy) defects in the pinned vortex lattice with a surplus (deficit) flux of $\frac12\Phi_0$. The interstitials can then be manipulated by a magnetic tip.~\cite{StrHofAus08a}

What is the response of the 2DEG system to a flux defect in the superconducting vortex lattice? To answer this question we will make two working assumptions. The first one is that the 2DEG is indeed in the integer quantum Hall state. In particular, we assume that the spatial variation in the field does not destroy this state. Second, we assume that the energetics of the system are dominated by the superconducting film. Importantly, this means that the quantization of flux by Cooper pairs remains valid and exact. We shall discuss the conditions under which these assumptions hold true in Sec.~\ref{sec:conc}. Consequently, the question is reduced to: What is the response of the integer quantum Hall state in 2DEG to a surplus/deficit flux of $\frac12\Phi_0$?

The answer is found by a thought experiment in which we slowly turn on the extra flux $\Phi(t)$ in time $t$ from zero to $\eta\Phi_0$, $\eta=\pm\frac12$. The total charge accumulated at the position of the defect is
\begin{equation}\label{eq:dQje}
\delta Q = - \int\d t\oint_C \vex j_e\cdot\vex n\;\d l,
\end{equation}
where $C$ is a contour in the plane containing the defect, $\vex n$ is a unit vector normal to it, and $\d l$ an element of length of $C$. In this process an electric field is induced in the plane by Faraday's law, $\oint_C \vex E\cdot\d\vex l = -\d{\Phi}/{c\d t}$. An electric field in the quantum Hall state with filling factor $\nu$ results in a transverse electric current $\vex j_e=\sigma_H\vex E\times\hat{z}$, where $\hat z$ is the normal to the plane and $\sigma_H=\nu e^2/h$ is the quantized Hall conductance. Altogether, we find from Eq.~(\ref{eq:dQje}),
\begin{equation}\label{eq:dQgen}
\delta Q = \eta\nu e.
\end{equation}
We note that for this result to be valid, the size of $C$ needs to be much larger than both the size of the flux defect, and the size of the smallest Landau orbit. The former is determined by the penetration depth, $\lambda_L$, in the superconducting film, which comes in through the use of Farady's law. We shall see its role more clearly in Secs.~\ref{sec:contAL} and~\ref{sec:eff}. The latter is of the order of the magnetic length $\ell_B=\sqrt{\hbar c/eB}$ and is implied by the physics of the quantum Hall state, as will be discussed in Sec.~\ref{sec:cont}.

Thus, at $\nu=1$, we have uncovered a bound state of charge $q=\eta e$ and flux $\Phi=\eta\Phi_0$. This is an almost literal realization of Wilczek's model of an anyon.~\cite{Wil82b} However, based on an analogy with the FQH quasiparticles, we expect a statistical angle $\theta=\eta^2\pi=\pi/4$, as opposed to the $2\eta^2\pi=\pi/2$ that would follow from Aharonov-Bohm phases. That this is the correct result can be affirmed by using the fusion rule\cite{wilzcekbook}: the statistical angle of a composite of $n$ anyons with statistical angle $\theta$ is $n^2\theta$. Putting two bound states $(q,\Phi)$ together we must obtain an electron, a fermion: $4\theta\equiv \pi \mod 2\pi$. This is consistent with $\theta=\pi/4$. In effect, the anyon carries the memory of its fermionic past.

These general findings will be confirmed by our detailed studies in the rest of the paper.

\section{Continuum model}\label{sec:cont}

As a first concrete model of our system, we will study a continuum model of noninteracting electrons in the 2DEG layer in magnetic field at filling factor $\nu=1$. In a uniform magnetic field the energy levels are organized in Landau levels with a degeneracy $N_\Phi=\Phi/\Phi_0$ per spin, the total number of flux quanta in the plane. The ultimate goal of this study is to find the levels in the periodic magnetic field generated by the Abrikosov lattice of vortices in the superconducting film and in the presence of vacancies and interstitial defects. Not surprisingly, the full solution can only be obtained numerically, but we will find analytical solutions for simpler cases. Especially, we will use the general solution for the ground states of a 2DEG in an \emph{arbitrary} magnetic field, due to Aharonov and Casher,~\cite{AhaCas79a} in the special case of ``Pauli'' electrons, i.e. free electrons with the gyromagnetic ratio $g=2$.

\subsection{Spin-polarized electrons}\label{sec:contpol}

We do not know the general form of the wavefunctions for other values of $0<g\neq2$, but for large enough $g\gg1$, the Zeeman coupling can be replaced with the condition that all electrons are in a single spin state, aligned with the magnetic field. We can then find the single-particle spectrum in the simple case of a uniform background field with vacancy and interstitial defects modeled by a $\delta$-function profile. The single-particle Hamiltonian is
\begin{equation}\label{eq:SchrH}
\mathcal{H}_{\mathrm{pol}} = \frac1{2m_e}\left(\vex p - \frac ec \vex A\right)^2,
\end{equation}
where $\vex p$ and $\vex A$ are the momentum and the vector potential operators.

We shall work in the dimensionless polar coordinate $(r\cos\varphi,r\sin\varphi)=\vex x/\sqrt2\ell_B$.  The magnetic field is $\vex B(\vex x) = \left(B_0 + \eta\Phi_0 \delta(\vex x)\right)\hat z$. The vector potential is given, in the symmetric gauge, by
\begin{equation}\label{eq:dAsymm}
\vex A = \frac1{\sqrt2}\frac{\Phi_0}{2\pi\ell_B}\left(r + \frac{\eta}{r}\right) \hat\varphi,
\end{equation}
where $\eta=\pm\frac12$ is the fraction of the flux quantum carried by the defect.

The Schr\"odinger equation $\mathcal{H}\Psi=E\Psi$ is solved in Appendix~\ref{app:Schr}. From Eq.~(\ref{eq:LLLpsi}), the states in the lowest Landau level (LLL) are given by
\begin{equation}
\Psi_l^{(\eta)}(z)\propto|z|^{-\eta}z^le^{-\frac12{|z|^2}},
\end{equation}
with the complex coordinate $z=re^{\ii\varphi}$. When $\eta=0$, the LLL contains $N_e=N_\Phi$ states. For a vacancy, $\eta<0$, the orbital quantum number $k=0$, and the angular momentum $l=0,1,\cdots$. All the states are pushed away from the center, so we still have $N_e=N_\Phi$ states in the LLL and a charge deficit at the center. For an interstitial, $\eta>0$, on the other hand, $k=0$, and $l=1,2,\cdots$. All the states are pulled in toward the center, so we lose the innermost state with $l=0$, whose energy is pushed up into the gap. Therefore, the LLL now contains $N_e=N_\Phi-1$ states.
The many-body ground state of the filled LLL is given by
\begin{equation}\label{eq:WFsymm}
\Psi^{(\eta)}\left({z_i}\right) \propto \prod_{i=1}^{N_e}|z_i|^{-\eta}\prod_{i>j}(z_i-z_j)e^{-\frac12\sum_i|z_i|^2}.
\end{equation}

For further use, we also note a different form of this wavefunction in the ``string gauge'' where the vector potential of the additional $\delta$-function flux is given by $\delta\vex A=(\sqrt2\ell_B r)^{-1}\eta\Phi_0\delta(\varphi)\hat\varphi$. This gauge can be obtained from the symmetric gauge in Eq.~(\ref{eq:dAsymm}) by a gauge transformation $\vex A\to\vex A-\vex \nabla\Lambda$, with $\Lambda=(\eta\Phi_0/2\pi)\varphi$. The single-particle wave functions transform as $\Psi\to\Psi_s=\Psi\exp(-\ii\frac{2\pi}{\Phi_0}\Lambda)=\Psi e^{-\ii\eta\varphi}$. The effect of this gauge transformation is to send $|z_i|\to z_i$ in Eq.~(\ref{eq:WFsymm}).
\begin{equation}\label{eq:WFstring}
\Psi_s^{(\eta)}\left({z_i}\right) \propto \prod_{i=1}^Nz_i^{-\eta}\prod_{i>j}(z_i-z_j)e^{-\frac12\sum_i|z_i|^2}.
\end{equation}

\subsubsection{Fractional charge}
Using results from the previous section we have for the charge density in the thermodynamic limit
\begin{eqnarray}
\rho_\eta(r) 	&=& e\left< \Psi^{(\eta)}\right|\sum_i\delta(\vex x -\vex x_i)\left|\Psi^{(\eta)}\right>\\
		&=& e\sum_l |\Psi_l^{(\eta)}(\vex x)|^2 \label{eq:rhoDef} = \frac{e}{2\pi\ell_B^2}F_\eta\left(r^2\right),
\end{eqnarray}
where,
\begin{equation}
F_\eta(\tau) = e^{-\tau}\sum_{l=0}^\infty\frac{\tau^{|l-\eta|}}{\Gamma(1+|l-\eta|)}.
\end{equation}
The charge displacement due to additional flux can be calculated as $\delta Q_\eta = Q_\eta-Q_0$, where $Q_\eta= \int \rho_\eta(\vex x) \d\vex x=e\int_0^\infty F(\tau)\d \tau$. Note that $F_0=1$.

For a vacancy,  $\eta<0$,
\begin{equation}
F_\eta(\tau)=1-\frac{\Gamma(-\eta,\tau)}{\Gamma(-\eta)},
\end{equation}
where $\Gamma(a,\tau)=\int_\tau^\infty s^{a-1}e^{-s}\d s$ is the incomplete gamma function. For an interstitial, $\eta>0$,
\begin{equation}
F_\eta(\tau)=1-\frac{\Gamma(1-\eta,\tau)}{\Gamma(1-\eta)}+\frac{\tau^\eta e^{-\tau}}{\Gamma(1+\eta)}.
\end{equation}
In both cases, $F_\eta(\tau)\to1$ for $\tau\gg1$, that is the length scale for the extra charge at the center is $\sim\ell_B$. Thus,
\begin{equation}
\delta Q_\eta = e\int_0^\infty \left[ F_\eta(\tau)-F_0(\tau) \right]\d \tau = \eta e,
\end{equation}
where we have used the identity $\int_0^\infty \Gamma(a,\tau)\d \tau=\Gamma(a+1)$. This confirms the value of the fractional charge we found by our general argument in Sec.~\ref{sec:sys} and clarifies the role of the magnetic length scale.

\subsubsection{Fractional statistics}
In order to find the statistics of the fractionally charged defects we calculate the Berry's phase,\cite{AroSchWil84a} defined as the phase accumulated by a state $\Psi_w$, as a parameter $w$ is taken around a contour $C$,
\begin{eqnarray}\label{eq:thB}
\theta_B &=& \ii\oint_C \left<\Psi_w\right|{\partial_w}\left|\Psi_w\right>\d w + \ii\oint_C \left<\Psi_w\right|{\partial_{\bar w}}\left|\Psi_w\right>\d \bar w \nonumber \\
	&\equiv& \theta_{B,w}+\theta_{B,\bar w}.
\end{eqnarray}
Here we have allowed for the possibility that $\Psi_w$ is not an entire function of $w$ (so it may depend on the complex conjugate $\bar w$, too). When $w$ parametrizes the encircling of one particle around the other, this is equal to twice the statistical angle. 

For a particle of charge $q$ encircling an area with magnetic flux $\Phi$, the Berry's phase is the same as the Aharonov-Bohm phase, $\theta_{AB}=2\pi (q/e)(\Phi/\Phi_0)$. This property can be used to confirm the value of the charge we obtained above. The ground state in the string gauge for an extra flux $\eta\Phi_0$ at a position with complex coordinate $w$ is given by Eq.~(\ref{eq:WFstring}) with the replacement $z_i^{-\eta}\to(z_i-w)^{-\eta}$. In a large system of size $L$ this is justified as long as $|w|\ll L/\ell_B$. A stronger argument will be given in the special case discussed in Sec.~\ref{sec:free}. Denoting the ground state when $\eta=0$ as $\Psi_0(z_i)$, we have
\begin{equation}\label{eq:dWF}
\Psi_w\equiv\Psi_s^{(\eta)}(z_i,w)\propto\prod_i (z_i-w)^{-\eta}\Psi_0(z_i).
\end{equation}
So, 
\begin{eqnarray}
\partial_w\Psi_w = -\eta\Psi_w\sum_i\frac1{w-z_i},
\end{eqnarray}
and $\partial_{\bar w}\Psi_w=0$. When $w$ is taken around a contour $C=\partial S$ this gives
\begin{eqnarray}
\theta_B &=& -\ii\eta\int \d^2 z\oint_C \frac{\left<\Psi_w\right|\sum_i\delta(z-z_i)\left|\Psi_w \right>}{w-z}\d w \nonumber\\
	&=& -\frac{\ii\eta}e \int \d^2 z \oint_C \frac{\rho_\eta(z-w)}{w-z}\d w 
\end{eqnarray}	
where we have used the definition~(\ref{eq:rhoDef}) for the charge density. Writing $\rho_\eta=\rho_0+\delta\rho_\eta$ we observe that the second term's contribution vanishes as $R^{-2}$ with the size $R$ of contour $C$ (in units of $\ell_B$). For the first term's contribution, we note that the contour integral evaluates to $2\pi\ii$ when $z\in S$ and 0 otherwise, thus 
\begin{equation}
\theta_B=2\pi\eta\frac{\Phi_S}{\Phi_0} + O\left(R^{-2}\right),
\end{equation}
where we have used the relationship between the charge and the magnetic flux in $S$, $\Phi_S$, for the $\nu=1$ integer quantum Hall state. Comparing with $\theta_{AB}$ we find the charge carried by the defects to be $\delta Q=\eta e$ as before.

The form of the wavefunction in~(\ref{eq:dWF}) suggests that we may write the wavefunction with two defects located at $w$ and $v$ as
\begin{equation}\label{eq:ddWF}
\Psi_{w,v}(z_i)\propto\prod_i (z_i-w)^{-\eta}(z_i-v)^{-\eta}\Psi_0(z_i).
\end{equation}
We will justify this form more strongly in the next section. With this choice we may now compute the Berry's phase again, when $w$ is taken around a contour $C=\partial S$ encircling $v$. The algebra is completely analogous to the previous case and, when $R-|v|\gg1$, we find,
\begin{equation}
\theta_B = \frac{2\pi\eta}e\int_{S}\rho_\eta(z)\d^2 z,
\end{equation}
where the charge $\eta e$ of the defect at $v$ must now also be taken into account in $\rho_\eta$. This gives $\theta_B=2\pi\eta(\Phi_S/\Phi_0+\eta)$ with the additional phase $2\theta$,
\begin{equation}
\theta = \eta^2\pi,
\end{equation}
being twice the statistical angle, as expected.

\subsection{Pauli electrons}\label{sec:free}

Electrons carry a magnetic moment 
\begin{equation}
\boldsymbol{\mu}=g \frac{e\hbar}{m_e c}\frac12\boldsymbol{\sigma},
\end{equation}
 where $\boldsymbol{\sigma}$ is a vector of Pauli matrices, and couple to the magnetic field by a Zeeman interaction term, $-\boldsymbol{\mu}\cdot\vex B$. For free electrons the gyromagnetic ratio $g=2$ with a very high precision. In a solid this value is renormalized and can be much higher. When $g=2$ there is a powerful method due to Aharonov and Casher,~\cite{AhaCas79a} which gives the ground states of the free Pauli electrons moving in a general profile of a perpendicular magnetic field.

The Hamiltonian for electrons in a perpendicular magnetic field is
\begin{equation}\label{eq:freeH}
\mathcal{H}_{\mathrm{free}}=\frac1{2m_e}\left( \vex p -\frac ec\vex A \right)^2 - \mu_z B.
\end{equation}
The ground states can be found for an arbitrary space-dependent applied magnetic field $B(\vex x)$ as $\psi_0(z)=f(z)e^{-\phi}$, where $f(z)$ is an entire function of the (dimensionless) complex coordinate $z$ and $\nabla^2\phi=2\pi B/\Phi_0$. For the sake of completness the derivation of this result is outlined in Appendix~\ref{app:AC}.

We will now add a flux defect with a general profile to the uniform background, $B_0$, given by $\delta B(\vex x)=\eta\Phi_0\delta_\xi(\vex x)$ where $\delta_\xi$ is a general function that vanishes over a characteristic length $\xi$ (in units of $\ell_B$) and with unit weight $\int\delta_\xi(\vex x)\d^2\vex x =1.$ Then, switching to the dimensionless complex coordinate representation, we may write $\phi(z)=\frac12|z|^2+\eta \phi_\xi(z)$ where
\begin{equation}\label{eq:Lphixi}
 \nabla^2 \phi_\xi(z) = 2\pi\delta_\xi(z),
\end{equation}
with the solution given, using the Laplacian Green's function $(2\pi)^{-1}\ln(r)$, as
\begin{equation}\label{eq:gxiGF}
\phi_\xi(z)=\int \ln|z-v|\delta_\xi(v) \d^2 v.
\end{equation}
 For illustration, we note that when we take the limit $\xi\to0$ we must have $\delta_\xi(z)\to\delta(z)$ and $\phi_\xi(z)\to\ln|z|$ which reproduces the results of the previous section. This provides further justification for Eqs.~(\ref{eq:dWF}) and~(\ref{eq:ddWF}).
 
Transforming to the string gauge, the many-body ground state with a defect at position $w$ is then
\begin{equation}
 \Psi_w(z_i)\propto\prod_i \frac{(z_i-w)^{-\eta}}{|z_i-w|^{-\eta}} e^{-\eta \phi_\xi(z_i-w)} \Psi_0(z_i),
\end{equation}
where $\Psi_0$ refers to the ground state in the absence of the defect ($\eta=0$) as before. Thus,
\begin{eqnarray}\label{eq:partw}
\partial_w\Psi_w &=& -\eta\Psi_w\sum_i\left[\partial_w \phi_\xi(z_i-w)+\frac12\frac1{w-z_i}\right], \\ \label{eq:partwbar}
\partial_{\bar w}\Psi_w &=& -\eta\Psi_w\sum_i\left[\partial_{\bar w} \phi_\xi(z_i-w)-\frac12\frac1{\bar w-z_i}\right].
\end{eqnarray}
Using the expression~(\ref{eq:gxiGF}):
\begin{equation}
\partial_w \phi_\xi(z-w)=\frac12\int\frac{\delta_\xi(u)}{u-(z-w)}\d^2 u.
\end{equation}
Similarly, $\partial_{\bar w}\phi_\xi$ is found by $w\to\bar w$.
Hence, following similar manipulations as in the previous section, the Berry's phase for transporting the defect around a contour $C=\partial S$ is found to be
\begin{equation}
\theta_{B,w}=-\frac{\ii\eta}{2e}\int\d^2 z\d^2 u\oint_C\frac{\rho_{\eta}(z-w;\xi)\left[\delta_\xi(u)+\delta(u)\right]}{w-(z-u)} \d w.
\end{equation}
Note that the second term in the square bracket is exactly half of the result found in the previous section for $\xi=0$. Similarly, $\theta_{B,\bar w}$ is found by $w\to \bar w$ and with $[\delta_\xi(u)-\delta(u)]$ in the numerator.

Writing $\rho_\eta=\rho_0+\delta\rho_\eta$ we see again that the correction from the second term in $\theta_{B,w}$ is $O(\max[\xi,1]/R)^2$, whereas the first term contributes exactly as before since $\int\d^2 u\delta_\xi(u)=\int\d^2 u\delta(u)=1$. The latter also means that the leading contribution to $\theta_{B,\bar w}$ vanishes. Thus,
\begin{equation}
\theta_B=2\pi\eta\frac{\Phi_S}{\Phi_0}+O\left(\frac{\max[\xi,1]}{R}\right)^2,
\end{equation}
confirming the value of fractional charge, $\eta e$, as well as illustrating the role of the size of the defect, $\xi$.

The statistical angle is calculated similarly from $\theta_B$ by taking a defect at $w$ around another defect fixed at $v$. The many-body wavefunction in the string gauge is given by
\begin{equation}
\Psi_{w,v}(z_i)\propto\prod_i\frac{(z_i-w)^{-\eta}}{|z_i-w|^{-\eta}}e^{-\eta \phi_\xi(z_i-w)}\Psi_v(z_i).
\end{equation}
The algebra is the same as before, but now $\rho_\eta$ contains the charge of the defect at $v$ so the leading contribution, when $R-|v|\gg\max[\xi,1]$, is given by $2\pi\eta\Phi_S/\Phi_0+2\theta$,
\begin{equation}
\theta=\eta^2\pi.
\end{equation}

\begin{figure}[t]
\begin{center}
\includegraphics[width=3.45in]{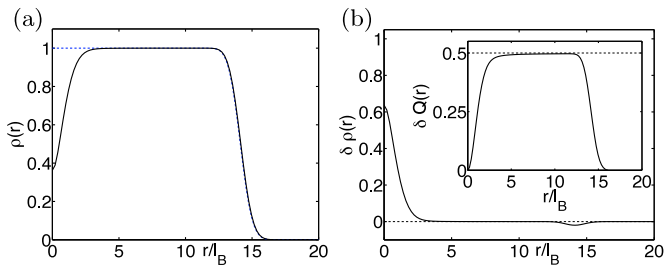}
\caption{(a) (color online) Charge density in units of $e/2\pi\ell_B^2$ in a system with $N_e=100$ electrons without (dashed blue line) and  with (solid black line) a defect ($\eta=-\frac12$) of the widened flux tube with $\xi=0.5\ell_B$. (b) The difference in density $-\delta\rho_\eta=\rho_0-\rho_\eta$. The inset shows 
the accumulated integrated charge $\delta Q(r) = -\int \delta\rho_\eta\d\vex x$ in units of $e$.}
\label{fig:widetube}
\end{center}
\end{figure}

As an example, let us consider a widened flux tube, with a size $\xi$ in units of $\sqrt2\ell_B$, given by 
\begin{equation}
\delta_\xi (z)=\frac 1\pi \frac{\xi^2}{\left(|z|^2+\xi^2\right)^2},
\end{equation}
which correctly tends to $\delta(z)$ as $\xi\to 0$. Then,
\begin{equation}
\phi_\xi(z) = \ln\sqrt{|z|^2+\xi^2},
\end{equation}
which can be obtained directly from Eq.~(\ref{eq:Lphixi}). Thus the many-body wavefunction in the string gauge for a flux tube at $w$ is found to be
\begin{equation}
\Psi_w \propto \prod_i (z_i-w)^{-\eta}\left( 1 + \frac{\xi^2}{|z_i-w|^2} \right)^{-\frac\eta2} \Psi_0.
\end{equation}
The derivatives needed for the Berry's phase calculation  are
\begin{equation}
\partial_w\Psi_w = -\frac\eta2\Psi_w\sum_i\partial_w\left[\ln\left(|z_i-w|^2+\xi^2\right)+\ln(z_i-w)\right],
\end{equation}
and similarly for $\partial_{\bar w}\Psi_w$. These are the same as Eqs.~(\ref{eq:partw}) and~(\ref{eq:partwbar}). Finally, the Berry's phase is
\begin{equation}\label{eq:wideBerry}
\theta_B = -\ii\frac\eta{2e}\int\d^2 z\oint_C \d w\left[\frac{\rho_\eta(z-w)}{w-z_\xi} + \frac{\rho_\eta(z-w)}{w-z}\right],
\end{equation}
where $z_\xi\equiv z-\xi^2/(\bar w-\bar z)$. Therefore, when $\xi\ll R$ the size of $C=\partial S$, both terms in Eq.~(\ref{eq:wideBerry}) have the same contribution, $2\pi\ii\int_S\d^2 z\rho_\eta = 2\pi\ii e\Phi_S/\Phi_0+O(1/R^2)$ to the integrals and we obtain $\theta_B = 2\pi\eta \Phi_S/\Phi_0$. The Berry's phase resulting from transporting one flux tube around the other can be calculated similarly to be $2\pi\eta(\Phi_S/\Phi_0+\eta)$ as long as $\max[\xi,1]\ll R$.

We have calculated the charge density and the accumulated charge in the vicinity of the widened flux tube numerically. The results are summarized in Fig.~\ref{fig:widetube} which once again confirms that the accumulated charge is exactly $\frac12 e$ and is drawn from the edge of the system.

\subsection{Pauli electrons in Abrikosov lattice}\label{sec:contAL}

We will now use this general method to study the more realistic situation of defects in a pinned Abrikosov lattice, which we take to be square.  Vortices are then separated by $\sqrt\pi\ell_B$, so that the flux through a unit cell (containing a vortex) is $\frac12\Phi_0$. The average magnetic field is $B_0=(\Phi_0/2\pi\ell_B^2)(1-\rho_d)$, where $\rho_d$ is the density of defects. The total field is obtained from the London equation
\begin{equation}\label{eq:L}
\lambda_L^2 \nabla^2 B - B = -\frac12\Phi_0 \sum_i \Delta_{\xi_0}(\vex x-\vex x_i),
\end{equation}
where $\vex x_i$ denotes the position of vortices, $\xi_0$ is the coherence length, $\lambda_L$ is the penetration depth, and $\Delta_{\xi_0}$ is the profile of a single vortex in the plane. In the pure London model this would be a delta function resulting from the phase singularity at the vortex center. A more realistic approach takes into account the finite vortex core size $\xi_0$ which leads to a broadening of the delta function. The form that is easy to implement in calculations and also gives good agreement with the experimentally observed field distribution is given by a Gaussian~\cite{Brandt}
\begin{equation}
\Delta_{\xi_0}(\vex x) = \frac1{2\pi\xi_0^2} \exp{\left(-\frac{|\vex x|^2}{2\xi_0^2}\right)}.
\end{equation}
If we let $\xi_0\to 0$ we recover the $\delta$-function vortices studied in Sec.~\ref{sec:contpol}.

\begin{figure}[t]
\begin{center}
\includegraphics[width=3.45in]{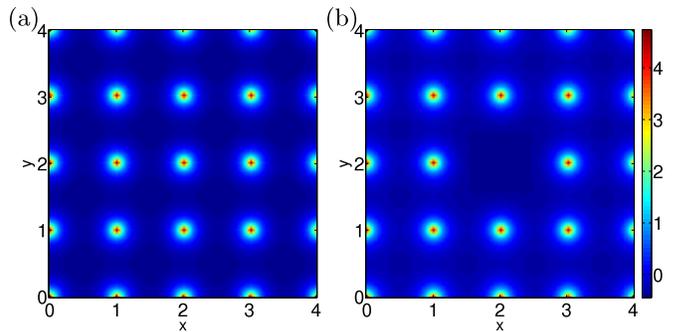}
\caption{(color online) Magnetic field (in units of $\Phi_0/\pi\ell_B^2$) of the pinned Abrikosov lattice, (a) without, and (b) with a vacancy defect. The plots are a partial view of the larger system with 64 pinning sites, $\lambda_L=0.2$ and $\xi_0=0.01$ in units of vortex separation, $\sqrt\pi\ell_B$.}
\label{fig:ACmag}
\end{center}
\end{figure}

\begin{figure}[h]
\begin{center}
\includegraphics[width=3.45in]{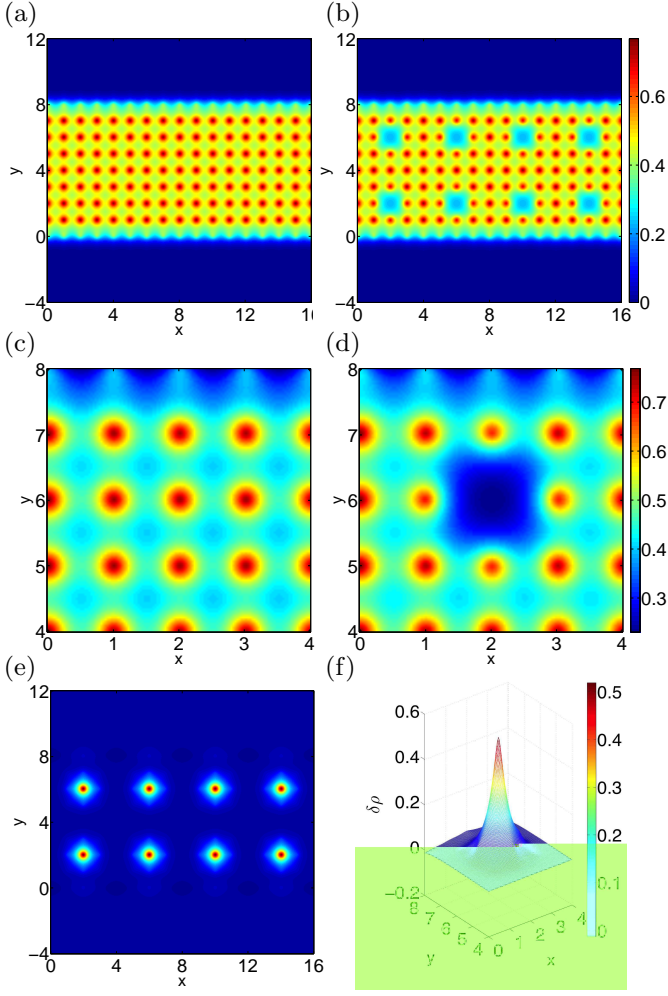}
\caption{(color online) The electronic density (in units of $e$) in the Abrikosov lattice (a, c) without, and (b, d) with defects. Panels (c) and (d) are zooms on (a) and (b) where a vacancy is introduced. The difference between (a) and (b) is shown in (e), and between (c) and (d) in (f). The system has 64 pinning sites, $\lambda_L=0.2$ and $\xi_0=0.01$ in units of vortex separation, $\sqrt\pi\ell_B$.}
\label{fig:ACrho}
\end{center}
\end{figure}

We decompose the field $B=B_0+\delta B+\delta B_d$, where $\delta B$ is the periodic modulation by the Abrikosov lattice and $\delta B_d$ is the field produced by defects. We choose to work in the Landau gauge where $\phi=y^2/2\ell_B^2$ from Eq.~(\ref{eq:phiL}). We have returned to the dimensionful quantities for clarity. This corresponds to periodic boundary conditions in the $x$ direction. Accordingly we choose the complete set of entire functions, $f_l(z)=\exp[-2\pi\ii l (x+\ii y)/L_x]$, where $l\in\mathbb{N}$ and $L_x$ is the size of the system in the $x$ direction. The (un-normalized) single-particle ground-state wavefunctions are then given by
\begin{equation}\label{eq:lattGS}
\Psi_l'(\vex x)=e^{\frac{2\pi}{L_x}l(y-\ii x)}e^{-\delta\phi}e^{-\delta\phi_d}e^{-\frac12\left({y}/{\ell_B}\right)^2},
\end{equation}
where $\nabla^2\delta\phi=\frac{2\pi}{\Phi_0}\delta B$ and $\nabla^2\delta\phi_d=\frac{2\pi}{\Phi_0}\delta B_d$.

Using the periodicity of the system we have $\delta B(\vex x) = \frac{\Phi_0}{2\pi\ell_B^2}\sum_{\vex G\neq0}\delta B_{\vex G} e^{\ii\vex G\cdot\vex x}$,
\begin{equation}\label{eq:dBG}
\delta B_{\vex G}=\frac{e^{-\frac12\xi_0^2 G^2}}{1+\lambda_L^2 G^2}.
\end{equation}
Here $\vex G$ is a reciprocal lattice vector. The $\vex G=0$ is excluded since the average extra flux $\int\delta B dS=0$. Fig.~\ref{fig:ACmag} shows the magnetic field, from Eq.~(\ref{eq:dBG}), in the pinned Abrikosov lattice when $\xi_0\ll\lambda_L\ll\ell_B$. This is the limit we are interested in: the first part ensuring that the superconductor is a strong type-II and the vortex core is small, and the second part ensuring that the vortices are well separated and defects are well localized. The result is a strongly modulated magnetic field. 

The Laplace equation~(\ref{eq:nabla2phi}) can be solved in the reciprocal space, $\delta\phi(\vex x)=\sum_{\vex G\neq 0}\delta\phi_{\vex G}e^{\ii\vex G\cdot\vex x}$,
\begin{equation}
\delta\phi_{\vex G}=-\frac1{\ell_B^2G^2}\delta B_{\vex G}=-\frac1{\ell_B^2G^2}\frac{e^{-\frac12\xi_0^2 G^2}}{1+\lambda_L^2 G^2}.
\end{equation}
We define a supercell of $N_d$ cells of the original lattice that contains one defect; thus, $\rho_d=1/N_d^2$. This superlattice  of defects has a corresponding reciprocal superlattice, whose vectors we denote by $\vex G_d$. Now, 
\begin{equation}
\delta\phi_d(\vex x)=\rho_d\sum_{\vex G_d\neq0}\delta\phi_{\vex G_d}e^{\ii\vex G_d\cdot\vex x}.
\end{equation}

Eigenfunctions $\Psi_l'(\vex x)$ in Eq.\ (\ref{eq:lattGS}) are linearly independent but in general do not form an orthonormal set. This complicates computation of observables. In the next step we thus map the states in~(\ref{eq:lattGS}) to an orthonormal basis by diagonalizing the overlap matrix
\begin{equation}
A_{lk}=\langle\Psi'_l|\Psi'_k\rangle.
\end{equation}
Denoting the orthogonal matrix that does so by $U$, $(U^\dag A U)_{lk}=a_l\delta_{lk}$ with eigenvalues $a_l$, the orthonormal basis is given by
\begin{equation}
\left|\Psi_l\right>=\frac1{\sqrt a_l}\sum_k U_{kl}\left|\Psi_k'\right>.
\end{equation}
Then, we may easily obtain the charge density in the many-body ground state,
\begin{equation}
\rho(\vex x)=e\sum_l|\Psi_l(\vex x)|^2.
\end{equation}

We have employed Fast Fourier Transform methods~\cite{DuhVet90a} to obtain the charge density in the system when defects are introduced in the vortex lattice. In Fig.~\ref{fig:ACrho} we show our results for the charge density. In the regular pinned vortex lattice, the electrons are mostly uniformly spread on the plane with regular peaks bound to the vortices. At $\nu=1$,  the average density is half an electron per vortex. A vacancy depletes the charge from around the vacant site, with an integrated charge deficiency which is exactly $\frac12 e$. This result allows us to conclude that a vortex defect binds exactly quantized fractional charge even when the periodic structure of the underlying magnetic field is taken into consideration.

\section{Lattice model}\label{sec:latt}

In this section we consider a lattice model to study the response of the 2DEG layer to the arrangement of vortices in the superconducting film. The continuum model in the previous section and the lattice model in this section can be thought of as the same system for different realizations of the gyromagnetic ratio. For large values of $g$ we can think of electrons as tightly bound to vortices.~\cite{BerRapJan05a} In this limit we can model the 2DEG with a tight-binding Hamiltonian for spin-polarized fermions on a square lattice where each plaquette is threaded with half a flux quantum, $\frac12\Phi_0$. 

The Hamiltonian is,
\begin{equation}\label{eq:lattH}
\mathcal{H}_{\mathrm{latt}}=\sum_{ij}t_{ij}e^{\ii\chi_{ij}}c_i^\dag c^{\vphantom{\dag}}_j + \sum_i \epsilon_i c_i^\dag c^{\vphantom{\dag}}_i,
\end{equation}
where $c_i$ annihilates an electron at site $\vex x_i$, $\epsilon_i$ is an onsite chemical potential, and $e^{\ii\chi_{ij}}$ are (gauge-dependent) Peierls factors incorporating the magnetic flux:
\begin{equation}
\chi_{ij}=\frac{2\pi}{\Phi_0}\int_{\vex x_i}^{\vex x_j} \vex A\cdot\d\vex x,
\end{equation}
with $\vex A$ the vector potential. The simplest model that produces the integer quantum Hall state is one with nearest and next-nearest neighbor hopping, $t$ and $t'$, respectively. The next-nearest neighbor hopping is needed to break the time-reversal symmetry. When $t'=0$ the smallest flux an electron sees is $\frac12\Phi_0$ through a plaquette, which is changed under time-reversal operation by $\Phi_0$. Therefore, the Hamiltonian is time-reversal invariant when $t'=0$ and the Hall conductance vanishes.

\begin{figure}[t]
\begin{center}
\includegraphics[width=3.45in]{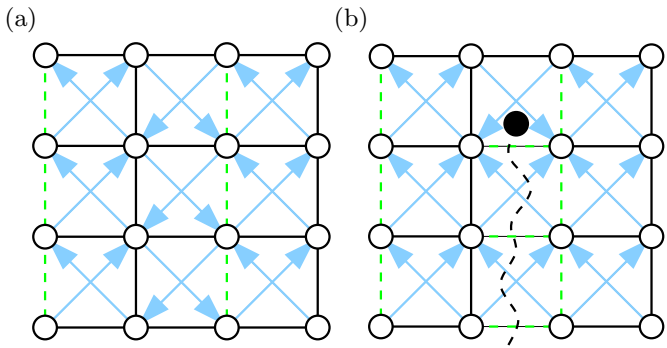}
\caption{(color online) (a) The Peierls factors, $e^{\ii\chi_{ij}}$, in the Landau gauge: (green) dashed links are $-1$, (black) solid links $+1$, and (blue) diagonal links $+\ii$ in the direction of the arrow. (b) The string gauge for a defect (black disk): every link that intersects the string (dashed wiggly line) acquires an extra $-1$.}
\label{fig:latt}
\end{center}
\end{figure}

\begin{figure}[t]
\begin{center}
\includegraphics[width=3.45in]{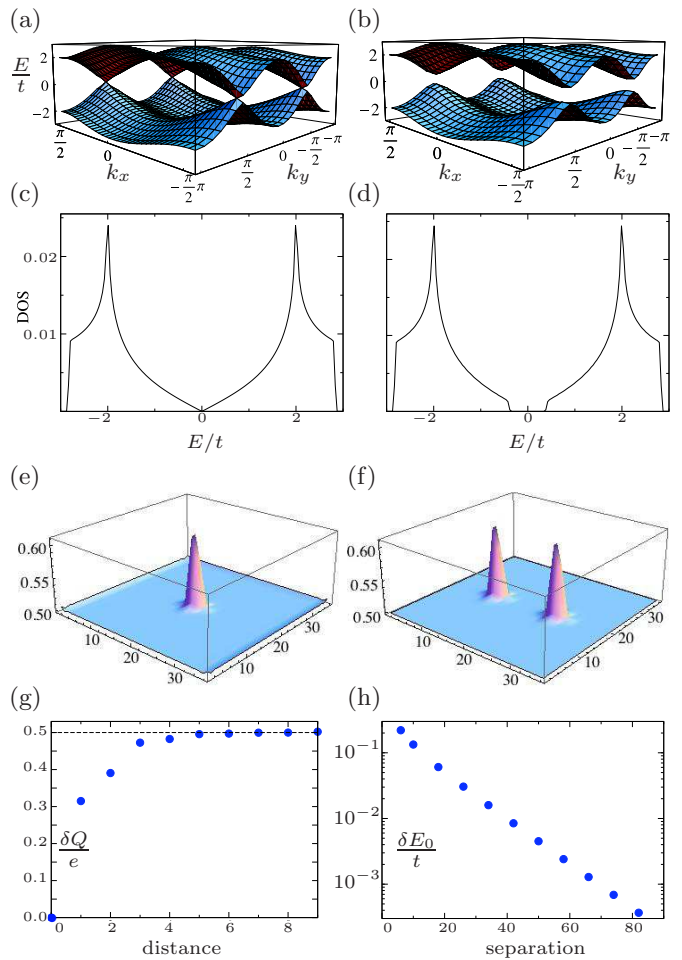}
\caption{(color online) The spectrum of the lattice model (a) without a gap, $t'=0$, and (b) with a gap $t'/t=0.1$. The density of states in the (c) gapless, and (d) gapped system. In all cases $\epsilon=0$ is assumed. Charge density with (e) a single defect, and (f) two defects in a $36\times36$ system and $t'/t=0.3$. (g) The integrated charge in a $40\times40$ system with open boundary conditions. (h) The energy splitting of zero modes of two defects vs. their separation in a $164\times10$ system with periodic boundary conditions.}
\label{fig:num}
\end{center}
\end{figure}

For a flux $\frac12\Phi_0$ uniformly distributed over a plaquette the vector potential in the Landau gauge is given by $\vex A=\frac{1}2\mathrm{sgn}(B)\Phi_0(0,x)$, where $B$ is the field perpendicular to the plane. The Peierls factors are then given by
\begin{eqnarray}
&&\chi_{i,i+\hat x}=0,\quad \chi_{i,i+\hat y}=\pi i_x, \\
&&\chi_{i,i+\hat x+\hat y}=\chi_{i+\hat x,i+\hat y}=\pi(i_x+\frac{1}2\mathrm{sgn}(B)).
\end{eqnarray}
These Peierls factors are pictured in Fig.~\ref{fig:latt}(a).
We may then choose a two-site unit cell $(i,i+\hat x)$ and form the spinor field $\psi_i=(c_i,c_{i+\hat x})^T$. For a uniform onsite potential $\epsilon_i=\epsilon$ the Hamiltonian can be written compactly in the reciprocal space, with the reduced Brillouin zone $\{\vex k: |k_x|\leq\pi/2,|k_y|\leq\pi\}$, as $\mathcal{H}_{\mathrm{latt}}=\sum_{\vex k} \psi^\dagger_{\vex k}\left(\epsilon-2tH_{\vex k}\right)\psi_{\vex k}$, where 
\begin{equation}\label{eq:lattHk}
H_{\vex k}=-h(\vex k)\sigma_z+\RR D(\vex k)\sigma_x+\II D(\vex k)\sigma_y.
\end{equation}
Here $\sigma_{x,y,z}$ are the Pauli matrices, and 
\begin{eqnarray}\label{hdk}
h(\vex k)&=&\cos k_y, \\ \nonumber
D(\vex k)&=&e^{\ii k_x}\left( \cos k_x-\ii m \sin k_x \sin k_y \right),
\end{eqnarray}
where we have introduced the ``mass'' $m=2\mathrm{sgn}(B)t'/t$. The energy spectrum is then given by
\begin{equation}
E(\vex k)=\epsilon \pm 2t\sqrt{\cos^2 k_x + \cos^2 k_y + m^2 \sin^2 k_x \sin^2 k_y},
\end{equation}
which is symmetric around $\epsilon$ and has a gap $8t'$ at the two independent ``nodes'' $\vex K^\pm=(\frac\pi2,\pm\frac\pi2)$. The symmetry of the spectrum is a general property of Eq.~(\ref{eq:lattHk}), since
\begin{equation}
\sigma_y H_{\vex k}^* \sigma_y = - H_{\vex k}.
\end{equation}
Therefore, $\sigma_y\psi_E^*$ is an eigenstate of energy $\epsilon-E$, if $\psi_E$ is one with energy $\epsilon+E$. The spectrum and the corresponding density of states are shown in Fig.~\ref{fig:num}(a-d).

A defect is introduced by an additional $\eta\Phi_0$ flux through one of the plaquettes. This will alter the Peierls factors by $\delta\chi_{ij}$, so that $\osum_{ij}\delta\chi_{ij}\equiv\pi \mod 2\pi$ around a closed loop containing the defect, and zero otherwise. We choose to work in the ``string gauge'' specified by a string originating from the defect and ending at a boundary: $\delta\chi_{\left<ij\right>}=2\pi\eta$ if the string cuts the bond $\vex x_j-\vex x_i$, and zero otherwise. This is shown in Fig.~\ref{fig:latt}(b). Two different choices of the string are related by a gauge transformation. It is important to note that our lattice formulation does not distinguish between an interstitial ($\eta=+\frac12)$ and  a vacancy ($\eta=-\frac12$), since the difference in flux is a full flux quantum through the smallest loop of the lattice (i.e. half a plaquette). As already mentioned above lattice electrons cannot distinguish such fluxes. 

\subsection{Hall conductance}

At half filling and for $t'\neq 0$ the lattice model exhibits precisely quantized Hall conductance $\sigma_H=\pm(e^2/h)$. To see this we employ the 
TKNN formula~\cite{tknn} which gives $\sigma_H=\mathcal{K}(e^2/h)$ in terms of the integer topological invariant
\begin{equation}\label{tknn1}
{\cal K}=\frac1{2\pi}\oint_{\partial\mathrm{BZ}}\d{\vex k}\cdot\mathcal{A}_{\vex k},
\end{equation}
with 
\begin{equation}\label{tknn12}
\mathcal{A}_{\bf k}={\sum_{s}}'\langle\psi_{s\vex k}| \ii\nabla_{\vex k}\psi_{s\vex k}\rangle.
\end{equation}
Here $\psi_{s\vex k}$ are eigenstates of $H_{\vex k}$ with band index $s$ and the sum is over occupied bands. 

The integral in Eq.~(\ref{tknn1}) represents the total Berry flux through the Brillouin zone. The largest contribution comes from the vicinity of the nodal points and it is easiest to evaluate $\mathcal{K}$ by first linearizing $H_{\vex k}$
near the nodes and then computing their contributions separately. As shown below in Sec.~\ref{sec:lowE} the linearized Hamiltonian near the node at $\vex K^\pm$ has the generic Dirac form
\begin{equation}\label{tknn2}
h_{\vex p}=p_x\sigma_x+p_y\sigma_y+m\sigma_z, 
\end{equation}
with the spectrum $E_{\vex p}=\pm\sqrt{\vex p^2+m^2}$ and ${\vex p}$ is the momentum relative to the nodal point. A quick analysis of Eqs.~(\ref{hdk}) reveals that the mass terms at the two nodes are equal. A straightforward calculation summarized in Appendix~\ref{app:tknn} yields 
\begin{equation}\label{tknn3}
\mathcal{K}^\pm=\frac12\mathrm{sgn}(m).
\end{equation}
Adding the two contributions, we thus find
\begin{equation}\label{tknn4}
\sigma_H={\rm sgn}(B){e^2\over h},
\end{equation}
as expected. This type of calculation is valid for any system with Dirac nodes. In graphene, for example, the mass terms created by a simple charge density wave have opposite sign and the Hall conductance vanishes, consistent with the fact that the system is time-reversal invariant in zero applied field.

\subsection{Exact diagonalization}

We have performed exact diagonalizations the lattice Hamiltonian~(\ref{eq:lattH}) on lattice sizes up to $50\times50$ in various settings. The system with a single defect and open boundary conditions supports a zero-energy bound state. Since the spectrum remains symmetric, standard arguments~\cite{SuSchHee80a} then show that the charge bound to the defect must be $\pm\frac12 e$ at half-filling depending on whether the zero mode is filled or empty. As seen in Fig.~\ref{fig:num}(g) the charge bound to a defect indeed integrates to $\frac12 e$ with a numerical precision within machine accuracy. The charge density profile in the ground state is shown in Fig.~\ref{fig:num}(e,f). The extra charge is localized around the defect center within a lengthscale $\sim m^{-1}$. For two defects, the two zero modes are slightly split due to tunneling between them with the energy splitting that decays exponentially with the defect separation. This splitting is plotted in Fig.~\ref{fig:num}(h). The effects of disorder in the on-site potential $\epsilon_i$ were discussed in Ref.~\onlinecite{WeeRosSer07a}. Specifically, we found that precisely quantized fractional charge persists even in the presence of disorder, as long as it is weak in comparison to the excitation gap.

\subsection{Lattice Berry's phase}

Since we have the complete spectrum of the lattice Hamiltonian, we can also calculate the Berry's phase accumulated by the ground state of the system with two defects as one is taken around the other.

Consider the ground state $|w\rangle$ with two defects, one placed at a fixed position and the other at $w$. As we take the second defect around the first one through the positions $w_1$, $w_2$, \dots, $w_N\equiv w_1$, we may calculate the accumulated phase at step $n$ though a generalized Bargamann invariant~\cite{Bar64a,SimMuk93a,RabArvMuk99a}
\begin{equation}\label{eq:lattB}
\theta_{B,n}=\arg \left<w_1|w_2\right>\left<w_2|w_3\right>\cdots
\left<w_{n-1}|w_n\right>\left<w_n|w_1\right>.
\end{equation}
In each step the phase of the overlap changes by an incremental amount; the phase of the product of overlaps is then equal to the sum of all such incremental changes. Of course $|w\rangle$ is defined up to an arbitrary phase. The product in~(\ref{eq:lattB}) is independent of this phase for all states.  Especially, the last overlap $\left<w_n|w_1\right>$ is included to make the product independent of the arbitrary phase of the initial and final states as well.

There is also a local gauge freedom in the Hamiltonian~(\ref{eq:lattH}) and thus in $|w\rangle$. Therefore, the overlaps $\left< w_r|w_{s}\right>$ in Eq.~(\ref{eq:lattB}) are gauge-dependent. This can be avoided by choosing to work with a gauge-invariant overlap,
\begin{equation}
{\langle w_r|w_{s} \rangle}_{\mathrm{inv}} =  \langle w_r| e^{-\ii\hat \chi_{0,rs}} |w_{s}\rangle,
\end{equation}
where $\chi_{0,rs}(\vex x_i)=\frac{2\pi c}{\Phi_0}\int_{ t_r}^{ t_s} A_0( t,\vex x_i)\d t$, $ t$ is time, and $A_0$ is the electromagnetic scalar potential in the chosen gauge. Basically, $\chi_{0,rs}(\vex x_i)$ is the temporal Peierls factor at site $\vex x_i$.
 
The significance of $\hat\chi_0$ can be understood by thinking about the dimensionless gauge-invariant flux $\Phi(C)=\oint_{C}\vec A\cdot\d\vec x$ for a temporal loop $C_{ij,rs}=( t_r,\vex x_i)\to( t_s,\vex x_i)\to( t_s,\vex x_j)\to( t_r,\vex x_j)\to( t_r,\vex x_i)$. 
On the one hand,
\begin{equation}\label{eq:PhiPt}
\Phi(C_{ij,rs})=\frac{\Phi_0}{2\pi}\left[\chi_{0,rs}(\vex x_j)-\chi_{0,rs}(\vex x_i)+\chi_{ij}( t_s)-\chi_{ij}( t_r)\right].
\end{equation}
On the other hand,
\begin{equation}\label{eq:Phitind}
\Phi(C_{ij,rs})=\int_{ t_r}^{ t_s}\d  t\int_{\vex x_i}^{\vex x_j} \vex E\cdot\d\vex x,
\end{equation}
where $\vex E=-\nabla A_0-\frac1c \partial_ t \vex A$ is the in-plane electric field. This is simply the total integrated electromotive force along the line connecting $\vex x_i$ to $\vex x_j$. For each flux quantum crossing $\vex x_j-\vex x_i$ in the time interval $ t_s- t_r$, we see from Eq.~(\ref{eq:Phitind}) that $\Phi(C_{ij,rs})$ changes by $\pm2\pi$, coming from the electric field  induced by the motion of the flux. Therefore, $\hat\chi_0$ in Eq.~(\ref{eq:PhiPt})  ensures that this induced field is correctly taken into account.

\begin{figure}[t]
\begin{center}
\includegraphics[width=2.1in]{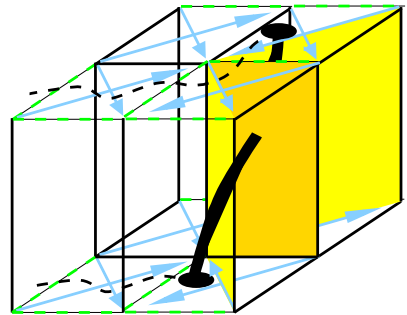}
\caption{(color online) The temporal string gauge. The vertical direction represents the time and style/color is as in Fig.~\ref{fig:latt}. As the defect is moved from a plaquette to the neighboring one, it threads the shaded temporal plaquettes, thus inducing an electric field. In the next time step the string is extended along the defect's path, which accounts for the induced electric field. The Peierls factors in the vertical direction remain equal to $+1$.}
\label{fig:temporal}
\end{center}
\end{figure}

We shall work in the temporal gauge where $A_0\equiv0$, where the gauge-invariant overlap is the same as the regular overlap. For a moving defect in the string gauge, this is satisfied when we extend the Dirac string exactly along the path of the defect. This can be seen by focusing on a primitive temporal plaquette, $\wp=C_{\left<ij\right>,\left<rs\right>}$. If the defect crosses the spatial side of $\wp$, $\Phi(\wp)=\pm\pi$. Since the Dirac string trails along the path of the defect, it also crosses the same spatial side of $\wp$, contributing exactly $\pm\pi$ to $\Phi(\wp)$ through the spatial Peierls factors in Eq.~(\ref{eq:PhiPt}). Thus, $\alpha_{0,\left<rs\right>}(\vex x_i)=\alpha_{0,\left<rs\right>}(\vex x_j)=0$. If the defect does not cross $\wp$, nor does the Dirac string. This is depicted in Fig.~\ref{fig:temporal}. The same conclusion can be drawn in the continuum formulation. In the string gauge $\vex A$ is perpendicular to the Dirac string, therefore in order that $A_0=\vex A\cdot\vex v/c=0$, the velocity, $\vex v$, of the defect must be tangent to the string, i.e. the Dirac string should be extended along the path of the defect.

\begin{figure}[t]
\begin{center}
\includegraphics[width=3.45in]{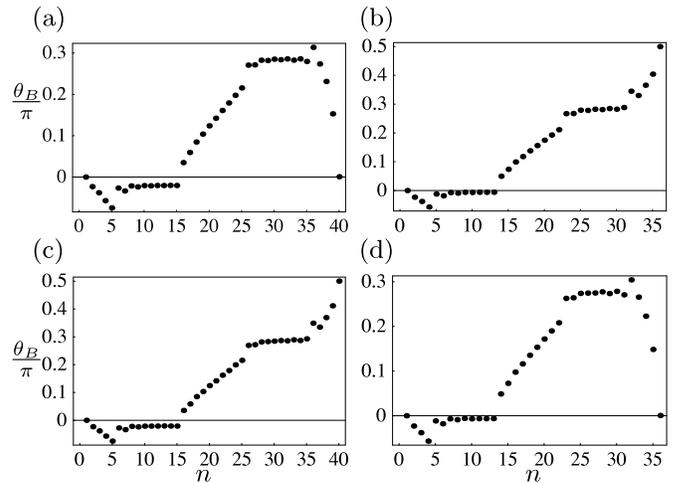}
\caption{The Berry's phase, $\theta_B/\pi$, and even-odd effect. In (a, b) there is a single defect in the system that is taken around a loop with (a) an even, or (b) an odd number of sites. In (c, d) the system contains two defects, of which one is taken around the other on a loop containing (c) an even, or (d) an odd number of sites.}
\label{fig:lattBerry}
\end{center}
\end{figure}

Fig.~\ref{fig:lattBerry} summarizes our numerical calculation of the Berry's phase. When a single defect is taken around a loop $C=\partial S$, the net effect can be understood as the Aharonov-Bohm phase of transporting the flux of the defect $|\eta|\Phi_0$ around the total charge inside the loop $\frac e2 N_S$, where $N_S$ is the number of sites enclosed by $C$. Hence, $\theta_B=\frac\pi2 (N_S \mod 2)$. This ``even-odd effect'' is seen in Figs.~\ref{fig:lattBerry}(a) and~\ref{fig:lattBerry}(b). When a second defect is introduced inside the loop $C$, the charge is reduced by $\frac e2$, which is equivalent in effect to reducing the number of sites by 1. Thus the ``even-odd effect'' must switch, as depicted in Figs.~\ref{fig:lattBerry}(c) and~\ref{fig:lattBerry}(d). The statistical angle can be inferred from the difference in the Berry's phase with and without the second defect, giving $2\theta=\pi/2$ as expected. 

\subsection{Low-energy limit}\label{sec:lowE}

Fractionl charge and statistics can also be found analytically by studying the Hamiltonian~(\ref{eq:lattH}) at low energies. This will give us a continuum description of the lattice model valid for the long-distance, collective behavior of the system. 

At half-filling the low-energy spectrum is dominated by quasiparticles around the two nodes $\vex K^\pm$. Expanding around the nodes, $\vex k=\vex K^\pm+\vex p$, the low-energy Hamiltonian is found to be $\mathcal{H}_{\mathrm{latt}}\to\sum_{\vex p}\Upsilon^\dag_{\vex p}H_{\vex p}\Upsilon_{\vex p}$, with $\Upsilon_{\vex p}=(\psi_{\vex K^++\vex p},\psi_{\vex K^-+\vex p})$, and
\begin{equation}
H_{\vex p} = \alpha_1 p_x + \alpha_2 p_y + \ii m\alpha_1\alpha_2,
\end{equation}
where $\boldsymbol{\alpha}=(\alpha_1,\alpha_2)=-(\id\otimes\sigma_y,\sigma_z\otimes\sigma_z)$. Since these matrices are diagonal in the nodal index, the two nodes are decoupled in the low-energy approximation. Moreover, by a rotation $\mathcal{S}=\exp({\ii\pi\id\otimes\sigma_z/4})\exp({\ii\pi\id\otimes\sigma_x/4})(\id\oplus\sigma_x)$ we find $H_{\vex p}\to\mathcal{S}^\dag H_{\vex p} \mathcal{S} = \id\otimes h_{\vex p}$, where $h_{\vex p}$ is given by Eq.~(\ref{tknn2}).

The effect of additional flux is to shift the momenta $\vex p \to \vex p - \frac ec\delta\vex A$, where $\delta\vex A$ is the vector potential of the additional field. For an additional flux $\eta\Phi_0$ in the symmetric gauge, $\delta\vex A = (\eta\Phi_0/2\pi r) \hat{\varphi}$. In real space, then,
\begin{equation}\label{eq:DiracHr}
H = \boldsymbol{\alpha}\cdot\left(\hat{\vex p}-\frac ec\delta\vex A\right) + \ii m\alpha_1\alpha_2, 
\end{equation}
where now $\hat{\vex p}=-\ii\hbar \nabla$ is the momentum operator.

The problem is reduced to Dirac fermions interacting with a (vanishingly thin) solenoid of flux $\eta\Phi_0$. It is known that the spectrum of this problem is not completely specified without extra boundary conditions at the origin.~\cite{Ger89a} Mathematically, this is related to having more than one self-adjoint extension of the Hamiltonian~(\ref{eq:DiracHr}). These self-adjoint extensions can be parametrized by a pair of angular variables $(\theta_{\mathrm{adj}}^+,\theta_{\mathrm{adj}}^-)$, one for each node. For $\pi/2<\theta_{\mathrm{adj}}^\pm<3\pi/2$ there is a bound state for the $\pm$ node with energy $|E|<m$ at any value of $\eta$. When $\theta_{\mathrm{adj}}^\pm=\pi$ the bound state at $\eta=\pm\frac12$ has zero energy. The bound state wavefunction in each case is square-integrable but diverges at the origin. We also note that without fixing $\theta_{\mathrm{adj}}^\pm$ one actually finds a continuum of bound states. This is due to the scale invariance of the Hamiltonian~(\ref{eq:DiracHr}), which means that by scaling the distance $r\to\lambda r$ we can go from a bound state $\Psi_{\mathrm{b}}(r)$ of energy $E$ to another one $\Psi_{\mathrm{b}}(\lambda r)$ with energy $E/\lambda$. This anomalous bound state spectrum collapses onto a single bound state energy by properly regularizing the problem at short distances. The self-adjoint parametrization provides such a regularization procedure.

The lattice itself provides a natural short-scale regularization, and our numerical results above show that indeed we have a single zero mode for each value of $|\eta|=\frac12$. Standard arguments~\cite{SuSchHee80a} then lead to a fractional charge $\delta Q=\eta e$. The sign is determined by noting~\cite{Ger89a} that the bound state energy tends to zero from negative (positive) values when $\eta$ is tuned from $0$ to $+\frac12$ ($-\frac12$) and therefore is to be included (excluded) in the ground state at half-filling.

To find the statistics we must go beyond the spectral properties of the low-energy theory to an effective field theory description. This can be done in a path-integral formulation where $\psi$ takes on the role of a dynamical field coupled to the electromagnetic gauge field $\delta\vec A=(\delta A_{0},\delta\vex A)$ of the defects. The mass term $\ii m\alpha_1\alpha_2$ breaks the time-reversal and parity symmetries in the Dirac Lagrangian. Integrating out fermionic field $\Upsilon$ at the one-loop order straightforwardly we obtain the effective action $S[\delta\vec A]=\int \d t\d\vex x \mathcal{L}_{\mathrm{CS}}+\cdots$, where
\begin{equation}\label{eq:LCS}
\mathcal{L}_{\mathrm{CS}}=\frac{\mathrm{sgn}(m)}{4\pi}\epsilon^{\alpha\beta\lambda}{\delta A}_\alpha\partial_\beta  \delta {A}_\lambda,
\end{equation}
is the Chern-Simons (CS) Lagrangian describing the topological sector of the effective theory and the dots indicate the remaining (non-topological part) of the theory. Of course, it is well-known that a CS term describes the topological aspects of QHE. The only difference here with the usual CS theory of QHE is that since the gauge field $\vec a_d$ describes the dynamics of defects in the (pinned) supercondcuting vortex lattice, it carries a magnetic flux quantized in units of $\frac12\Phi_0$ instead of $\Phi_0$.

The charge and statsitics of a defect carrying flux $\eta\Phi_0$ can now be calculated in the standard way~\cite{Wen04a} (see also Sec.~\ref{sec:eff}) from the CS theory, yielding the charge $\delta Q=\eta e$ and the statistical angle $\theta=\eta^2\pi$, as obtained previously.

\section{Effective theory}\label{sec:eff}

In this section we formulate an effective theory to describe the proposed system and the interaction between the 2DEG and the superconducting film. Using standard duality mappings, we show that the topological properties of the system in the vortex lattice phase are described by a Chern-Simons-Maxwell theory. Using this effective theory we derive the fractional charge and statistics of its excitations. 

Our notation in this section is as follows: the space-time vector is denoted by $\vec x=(x^0,x^1,x^2)$ with $\vex x=(x^1,x^2)$ the spatial and $x^0=c t$ the temporal parts. We use the relativistic notation in Minkowski space-time $\vec u\cdot\vec v=u_\mu v^\mu=g_{\mu\nu}u^\mu v^\nu$ and $(\vec\partial\times\vec u)^\mu=\epsilon^{\mu\nu\lambda}\partial_\nu u_\lambda$, where $g=\diag(-1,1,1)$ is the Minkowski metric.  

\subsection{Formulation}

The 2DEG at odd inverse filling factor $1/\nu$ is described by the Chern-Simons effective action~\cite{Wen04a}
\begin{equation}
S_{\mathrm{CS}}=\int\d^3x\left[-\frac1{4\pi\nu}\vec a\cdot(\vec \partial\times \vec a)+\frac{e}{2\pi} \vec a\cdot(\vec \partial\times \vec A) \right].
\end{equation}
We remind the reader that the topological gauge field $\vec a$ couples to the electromagnetic gauge field $\vec A$ in just the right way as to lock the electromagnetic field $\vec \partial\times \vec A$ to the electronic current $\vec j_e=\delta S_{\mathrm{CS}}/\delta \vec A=\frac{e}{2\pi}(\vec \partial\times \vec a)$:
\begin{equation}\label{eq:jnuB}
\frac{\delta S_{\mathrm{CS}}}{\delta \vec a}=0 \Rightarrow \vec j_e = \nu e\: \vec\partial \times \vec A.
\end{equation}
In particular, the electron density $j^0_e=\nu e B$.

We shall describe the superconducting film at zero temperature by the action $S_{\mathrm{pin}}+S_{\mathrm{XY}}$ where $S_{\mathrm{pin}}$ represents the effects of the lattice of pinning sites and $S_{\mathrm{XY}}$ is the 3DXY model, which we first write on the discretized Euclidean space-time,
\begin{equation}
S_{\mathrm{XY}} = -\kappa_0 \ell \sum\cos(\vec \Delta\vartheta- \ell\:e_*\vec A).
\end{equation}
Here, $\kappa_0$ is the superconducting phase stiffness, $\vartheta$ is the superconducting phase, $e_*=2e$ is the Cooper pair charge, $\vec \Delta$ is the lattice difference operator and the sum is over the points of a space-time cubic lattice with the spacing $\ell$ explicitly included to make the transition to the continuum limit transparent. This action describes the phase ordering of the superconducting transition in a magnetic field with the London penetration depth $\lambda_L=1/\sqrt{4\pi\kappa_0 e_*^2}$. In the continuum limit it could be obtained from the Ginzburg-Landau theory by taking the amplitude of the superconducting order parameter to be constant. Using the Villain approximation and standard duality mappings~\cite{Pes78a,NagLee00a} we may rewrite the 3DXY action in Minkowski space-time continuum in the ``current representation,''
\begin{equation}\label{eq:3DXYdual}
S_{\mathrm{XY}} = \int\d^3x \left[ -\frac1{2\kappa_0}(\vec \partial\times \vec s)^2-e_* (\vec \partial\times \vec A)\cdot\vec s+2\pi \vec j_v\cdot \vec s \right],
\end{equation}
where $s$ is the dual vortex field and $\vec j_v$ is the conserved vortex current, $\vec \partial\cdot \vec j_v=0$. 

In the following we shall take the main effect of $S_{\mathrm{pin}}$ to be the pinning of the background vortex lattice and to provide a one-body potential for the defects that preferentially positions them at the interstitial locations. Therefore we will drop this term from the dynamics with no change in topological properties.

\subsection{Phases of the system}
The 3DXY action~(\ref{eq:3DXYdual}) has two phases: (1) a vortex-lattice phase, where $\left<\vec s^2\right>=0$ in the dual or $\left< e^{\ii\vartheta} \right>\neq0$ in the original problem; (2) a vortex-liquid phase, where $\left<\vec s^2\right>\neq0$ in the dual or $\left< e^{\ii\vartheta}\right> =0 $. We will now derive the effective action of the combined system,
\begin{equation}
S=S_{\mathrm{CS}}+S_{\mathrm{XY}},
\end{equation}
in each of these two phases.

(1) In the vortex-lattice phase we may integrate out the dual vortex field $\vec s$ by noting that if we write the conserved current $\vec j_v=\frac1{2\pi}\vec \partial\times\vec \phi$ where $\vec \phi$ is a hydrodynamic flux variable, we have
\begin{equation}
S_{\mathrm{XY}}=\int\d^3x\left[ -\frac1{2\kappa_0}(\vec \partial\times \vec s)^2 + (\vec \phi-e_*\vec A)\cdot(\vec \partial\times \vec s) \right].
\end{equation}
Since $\left<\vec s^2\right>=0$ we may integrate $\vec s$ out to find
\begin{eqnarray}
S_{\mathrm{XY}}\to  \frac{\kappa_0}2\int\d^3x
(\vec \phi-e_*\vec A)_\perp^2,
\end{eqnarray}
where the subscript $\perp$ indicates that only the transverse parts of the fields enter this term. At this point we separate the static and dynamic parts of $\vec A$ and $\vex\phi$ coming from, respectively, the pinned background vortex lattice and the defects:
$\vec A=\vec A_0 + \delta\vec A$ and $\vec\phi=\vec\phi_0+\delta\vec\phi$. Then we may shift the electromagnetic gauge field of defects $\delta\vec A\to \delta\vec A+\delta\vec \phi/e_*$ and then integrate out $\delta\vec A$ to find the Chern-Simons-Maxwell form
\begin{equation}\label{eq:eff}
S_{\mathrm{eff}}=\int\d^3x\left[\frac{e}{e_*}\vec j_v\cdot \vec a -\frac1{4\pi\nu}\vec a\cdot(\vec \partial\times \vec a)-\frac1{2\kappa}(\vec \partial\times \vec a)^2 \right],
\end{equation}
where $\kappa=(2\pi e_*/e)^2\kappa_0$ and $\vec j_v$ is the \emph{total} vortex current of the pinned lattice and defects.
We shall study the topological properties of this action shortly.

(2) In the vortex-liquid phase, the dual field $\vec s$ acquires a mass and the Maxwell term is replaced with a mass term
\begin{eqnarray}
S_{\mathrm{XY}}&\to&\int\d^3x\left[\frac{\mu_0}2\vec s^2+ \vec s\cdot\vec \partial\times(\vec \phi-e_*\vec A)\right] \\
&\to& -\frac1{2\mu_0}\int\d^3x \left[\vec \partial\times(\vec \phi-e_*\vec A)\right]^2,
\end{eqnarray}
after integrating $\vec s$ out. Shifting and integrating out $\delta\vec A$ as before we now find a mass for $\vec a$:
\begin{equation}
S'_{\mathrm{eff}} = \int\d^3x \left[\frac{e}{e_*}\vec j_v\cdot \vec a-\frac1{4\pi\nu}\vec a\cdot(\vec \partial\times \vec a)+\frac\mu2\vec a_t^2 \right],
\end{equation}
where $\mu=(e/2\pi e_*)^2\mu_0$.
Thus in the vortex liquid phase the gauge field $\vec a$ is gapped and its topological effects are absent. 

\subsection{Fractional charge and statistics}

The fractional charge of defects in the vortex lattice phase, i.e. the Chern-Simons-Maxwell effective theory~(\ref{eq:eff}), is found as follows. From Eq.~(\ref{eq:eff}) we find $\delta S_{\mathrm{eff}}/\delta \vec a = 0$ gives
\begin{equation}
\vec j_e-\xi\vec \partial\times \vec j_e - |\eta|\nu e \vec j_v = 0,
\end{equation}
where $\vec j_e=\frac{e}{2\pi}\vec \partial\times \vec a$, $|\eta|\equiv e/e_*=\frac12$, and $\xi\equiv 2\pi\nu \kappa^{-1}$ is the coherence length. This modifies Eq.~(\ref{eq:jnuB}), reflecting the effect of the superconducting film. After a Fourier transform we find
\begin{eqnarray}\label{eq:jev}
\vec j_e(\vec k) = \frac{|\eta| \nu e}{1-\xi^2\vec k^2}\bigg[ \vec j_v +\ii \xi \vec k\times \vec j_v 
- \xi^2(\vec k\cdot \vec j_v)\vec k \bigg].
\end{eqnarray}
So, the excess charge density vanishes exponentially over a distance $\xi$, which is basically the size of the defect. The total charge $Q( t)=\int\d\vex x j^0( t,\vex x) = j^0_e( t,\vex k=0)$, which by Eq.~(\ref{eq:jev}) is
\begin{eqnarray}
Q( t)  &=& |\eta| \nu e j_v^0( t,\vex k=0) \nonumber \\
	&=& |\eta| \nu e \int \frac{\d^2 \vex x}{2\pi}(\vec\partial\times\vec\phi)^0,
\end{eqnarray}
and we have used the definition of the vortex current in terms of the flux variable $\vec\phi$. The integral here is nothing but the total vorticity (the winding number) of the lattice, a topological invariant. Introducing an interstitial (vacancy) at the origin changes the vorticity by $+1$ ($-1$). Thus, defining the sign of $\eta$ to be the same as the vorticity, the charge bound to a defect is found to be
\begin{equation}
\delta Q=\eta\nu e.
\end{equation}

The fractional statistics can be established by a Berry's phase calculation in the effective theory. To this end let's write the vortex current in Eq.~(\ref{eq:eff}) as $\vec j_v=\vec j_0+\delta\vec j_1+\delta\vec j_2$ where $\vec j_0$ is the vortex current of the background pinned lattice and $\delta j_{s}^{\mu}( t,\vex x)=(\d x_{s}^\mu/\d  t)v_s\delta(\vex x - \vex x_s( t))$ is the vortex current for defects $s=1,2$, with world-lines $\vec x_{s}=(c t,\vex x_s)$ and vorticity $v_s=\pm1$. We take a contour $\Gamma_2$ for $\vex x_2( t)$ that encircles the stationary position $\vex x_1=0$ and ask what is the Berry's phase contribution from the cross terms of $\delta \vec j_1$ and $\delta \vec j_2$.

By integrating out the Chern-Simons field we find 
\begin{equation}\label{eq:effv}
S_{\mathrm{eff}} \to \pi\nu|\eta|^2 \int \d^3 \vec k\; {j_{v}^{\mu}}(-\vec k)M^{-1}_{\mu\nu}(\vec k)j_{v}^{\nu}(\vec k),  
\end{equation}
where
\begin{equation}\label{eq:M}
M_{\mu\nu}=  \xi\left(k_\mu k_\nu-g_{\mu\nu} \vec k^2\right) + \ii\epsilon_{\mu\nu\alpha} k^\alpha.
\end{equation}
By substituting the vortex current terms in Eq.~(\ref{eq:effv}) we find the exchange Berry's phase
\begin{equation}
\theta_B = \pi\nu|\eta|^2 \int \d^3 \vec k \; \delta j_{1}^{\mu}(-\vec k) M^{-1}_{\mu\nu}(\vec k) \delta j_{2}^{\nu}(\vec k).
\end{equation}
In order to invert the matrix $M$ in Eq.~(\ref{eq:M}), we need to fix the gauge for $\vec a$. We do this by adding a term $M^{\mu}_{\nu}\to M^{\mu}_{\nu} + \zeta k^\mu k_\nu$. In general $\zeta$ can also be $\vec k$-dependent producing a nonlocal gauge fixing term.  
The result is
\begin{eqnarray}\label{eq:Mm1}
M^{-1}_{\mu\nu}= \frac{\xi}{1-\xi^2\vec k^2}\left[  g_{\mu\nu} +\frac\ii{\xi\vec k^2}\epsilon_{\mu\nu\alpha} k^\alpha \right] \nonumber \\
+ \left( \frac1{\zeta \vec k^2} -  \frac{\xi}{1-\xi^2\vec k^2} \right)\hat k_\mu \hat k_\nu.~~~~
\end{eqnarray}
By current conservation, $\vec \partial\cdot \vec j_{s}=0$, the term $\sim \hat k_\mu \hat k_\nu$ does not contribute to $\theta_B$, as is required by gauge invariance. After some tedious but ultimately standard calculation we find $\theta_B= \theta_{\mathrm{dyn}} + \theta_{\mathrm{top}}$, where the topological (dynamical) phase is contributed by the off-diagonal (diagonal) part of $M^{-1}$ in Eq.(\ref{eq:Mm1}). The topological phase is twice the statistical angle, $\theta_{\mathrm{top}}=2\theta$.

The dynamical phase
\begin{equation}
\theta_{\mathrm{dyn}} = \nu\eta_1\eta_2 \frac c\xi \int_0^T \d t K_0\left( \frac{|\vex x_2( t)|}{\xi}\right),
\end{equation}
where $\eta_s=|\eta|v_s$, $K_0$ is the modified Bessel function, and we have used the fact that
\begin{equation}
\int \frac{\d^2\vex q}{2\pi} \frac{e^{\ii\vex q \cdot \vex r }}{\vex q^2 + 1}=K_0(|\vex r|).
\end{equation}
It depends on the encircling time $T$  and the shape of the contour $\Gamma_2$. For a circular contour of radius $R$ the integral evaluates to $T K_0\left(R/\xi\right)$. For a large enough contour, then, $\theta_{\mathrm{dyn}}\to0$ exponentially over a distance $\xi$.

By contrast the topological phase
\begin{equation}
\theta_{\mathrm{top}} = \nu\eta_1\eta_2 \oint_{\Gamma_2}  \left[\frac{\xi}{|\vex x|} - K_1\left(\frac{|\vex x|}\xi\right) \right] \frac{\vex x \times \d\vex x}{\xi |\vex x|},
\end{equation}
does not depend on the encircling time $T$. For a circular contour of radius $R$, the integral evaluates to $2\pi[1 - (R/\xi) K_1(R/\xi)]$. So, for large $R$
\begin{equation}
\theta =  \pi\nu\eta_1\eta_2,
\end{equation}
with an exponentially vanishing correction over a distance $\xi$.

\section{Experimental signatures}\label{sec:exp}
Fractional charge has been unambiguously detected in FQH systems using a quantum antidot electrometer \cite{goldman1} and shot noise analysis.\cite{saminadayar1,picciotto1}
In other systems, such as the dimerized polyacetylene chain,\cite{{SuSchHee79a}} the experimental detection presents a greater challenge and the interpretation of the results is less straightforward.\cite{heeger1} Experimental detection of the fractional statistics in FQH systems to date remains elusive. Although the claim has been made to detect the (abelian) fractional statistics in a Laughlin state\cite{camino1} the interpretation of these results remains controversial. It appears difficult to disentangle the effects of fractional charge from the statistics, these being, in essence, two complementary manifestation of the same underlying many-body wavefunction. It has been pointed out\cite{sarma1,bonderson1} that the detection of the non-abelian exchange structure expected to occur in the Moore-Read pfaffian state could be in fact more straightforward, as this effect is not directly tied to fractional charge. In general, unambiguaus detection of the fractional statistics in {\em any} physical system remains an unsolved problem and a  challenge.

Against this backdrop we now discuss possible experimental signatures of the fractional charge and statistics in our proposed system.  We outline a concrete experimental setup for probing the fractional charge bound to the vortex defect and show how it can be measured \emph{directly} in a bulk transport measurement. In fact this measurement appears to us more straightforward than any other scheme for fractional charge detection discussed in the existing literature. This simplicity arises from the fact that the number of vortices traversing the width of the system can be counted precisely through the Josephson relation, a fundamental property of the superconductor. The charge bound to these vortices, on the other hand, can be measured accurately owing to the precisely quantized Hall conductance, a fundamental property of the 2DEG.

Similar considerations allow for controlled manipulation of vortex defects. This opens up a possibility of moving them, in principle at least, along any desired trajectory in the system. We describe below an idea for the anyon shuttle, an all-electric system for anyon manipulation. The ability to move anyons in a controlled fashion should aid the future experiments aimed at probing their exchange phase, although drawing on the experience with FQH systems we expect this to present a significant experimental challenge.

\subsection{Fractional charge}

\begin{figure}[t]
\begin{center}
\includegraphics[width=3in]{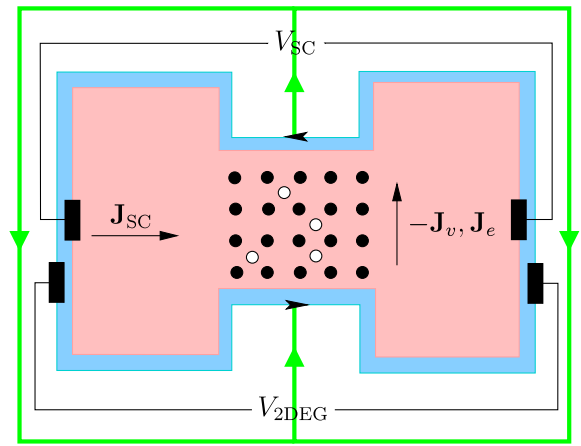}
\caption{(color online) Experimental setup for measuring the fractional charge. A supercurrent, $\vex J_{\mathrm{SC}}$, is induced in the superconducting film, which causes a vortex flow, $\vex J_v$, of unpinned defects (white circles) on the background of pinned vortex lattice (black circles) across the constriction along with a corresponding electric current, $\vex J_e$, in the 2DEG. A Hall voltage, $V_{\mathrm{2DEG}}$, is generated in the 2DEG and a voltage drop $V_{\mathrm{SC}}$ develops across the superconductor. This measures the fractional charge $\delta Q = (V_{\mathrm{2DEG}}/2V_{\mathrm{SC}})e$. The small arrows show the chiral edge current in the 2DEG and the thick (green) lines show the wires  that carry the electric current in 2DEG.}
\label{fig:exp}
\end{center}
\end{figure}

The setup geometry is shown in Fig.~\ref{fig:exp}. The system is subject to a perpendicular magnetic field $|\vex B|\gtrsim B_M$, the matching field, which produces weakly pinned vortices in interstitial positions. In the 2DEG a charge $\delta Q=e/2$ is bound to such defects.  A supercurrent, $\vex{J}_{\mathrm{SC}}$, is induced in the superconducting film producing a Magnus force, $\vex{F}_M = \vex{J}_{\mathrm{SC}}\times\vex B/c$, on the vortices. The current density is largest in the constriction. By suitably tuning the magnitude of the current one could arrange for the force only to affect the interstitial vortices, generating a vortex current $\vex{J}_v$ in the film along with an electric current $\vex{J}_e$ in the 2DEG as fractional charges bound to defects cross the constriction. The vortex and electric currents result in, respectively, a voltage drop, $V_{\mathrm{SC}}$, across the superconducting film and a Hall voltage, $V_{\mathrm{2DEG}}$, across the 2DEG. As we demonstrate below the ratio of these two voltages provides a direct measure of the fractional charge.

The superconducting phase difference, $\Delta\vartheta$, across the film has a time-dependence given by the Josephson relation
\begin{equation}
\frac{\d\Delta\vartheta}{\d t} = \frac{2e}{\hbar} V_{\mathrm{SC}}.
\end{equation}
When a vortex crosses the constriction it accounts for a change of $2\pi$ in the phase $\Delta\vartheta$, therefore
\begin{equation}\label{eq:Jv}
|\vex J_v|=\frac1{2\pi}\frac{\d\Delta\vartheta}{\d t} = \frac{2e}{h} V_{\mathrm{SC}}.
\end{equation}
The corresponding electric current in the 2DEG is simply $\vex J_e=\delta Q\;\vex J_v$. The Hall voltage is therefore found to be
\begin{equation}
V_{\mathrm{2DEG}}=\frac1{\sigma_H}|\vex J_e|= \frac{2\delta Q}{\nu e} V_{\mathrm{SC}},
\end{equation}
where we have used the quantized value $\sigma_H=\nu e^2/h$ in the quantum Hall state. So, we arrive at a particularly simple relation for the fractional charge of defects, namely,
\begin{equation}\label{eq:fracch}
\delta Q= \frac{V_{\mathrm{2DEG}}}{2V_{\mathrm{SC}}} \nu e.
\end{equation}
The fractional charge bound to the vortex is thus simply related to the ratio of two experimentally measurable voltages. It is worth emphasizing that Eq.\ (\ref{eq:fracch})  is an exact relation whose validity relies only on the fundamental properties of a superconductor and a 2DEG in the quantum Hall regime at filling fraction $\nu$.

\subsection{``Anyon shuttle'' and interference measurements}

\begin{figure}[t]
\begin{center}
\includegraphics[width=2.4in]{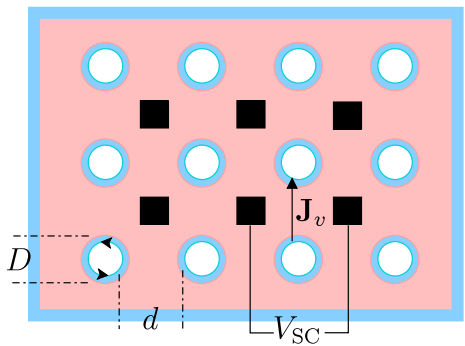}
\caption{(color online) The ``anyon shuttle.'' A superlattice of holes with diameter $D$ and separation $d$, both $\gg\lambda_L$, is used as source and sink of anyons. The arrows on the lower left hole shows the chiral edge currents concealed on other holes for clarity. The black squares are the leads on the superconducting film.}
\label{fig:shuttle}
\end{center}
\end{figure}

Observing fractional statistics is a much more difficult feat requiring precise interference experiments. However, in our system, the fact that the anyons have a physical flux attached to them might prove useful. For instance, it is possible to perform \emph{bulk}, as opposed to edge, measurements on the system. Bulk sources and drains of fractional particles may be created by building sources and sinks of flux tubes. Explicitly, we will in the following lay out an \emph{all electrical} scheme for shuttling fractional particles around along any desired path of a superlattice of holes drilled in both the layers of the system. A sketch of one such design is shown in Fig.~\ref{fig:shuttle}.

An array of holes drilled through both the superconducting film and the 2DEG layer provides the sources and drains of anyons through a supply of superconducting vortices trapped in the holes. The hole diameter, $D$, and spacing, $d$, between them are both much larger than the penetration depth $\lambda_L$. In equilibrium the magnetic flux through each hole will be quantized in units of $\Phi_0/2$. In the 2DEG each hole has a set of low-energy chiral  edge states with discrete spectrum due to the finite hole size. These edge states can accommodate extra charge which is crucial for the functioning of the anyon shuttle.

Now imagine we attach wires to the superconducting film at the centers of the square leads as indicated in Fig~\ref{fig:shuttle}. Through these wires we can feed appropriate supercurrents. When the current is sufficiently strong the resulting Magnus force $\vex{F}_M$ moves the vortices. If the geometry is right vortices will move from one hole to another, carrying fractional charge with them. As before, the number of vortices traversing between the holes can be determined from the voltage drop via the Josephson relation, Eq.~(\ref{eq:Jv}). Using a suitable sequence of current pulses it is possible to shuttle vortices along any desired path in the device, effectively braiding them as required in various quantum computation schemes. An important feature of this scheme is its being completely electrical, foregoing the need for any moving parts. We believe this is a major advantage especially in view of possible applications of non-Abelian  anyons in quantum computation as briefly discussed in Sec.~\ref{sec:conc}.

Using this shuttling scheme it should also be possible to perform bulk interference measurements to measure the fractional statistics of the anyons. A variety of such measurements exists, especially in the context of FQH physics. Assessing the merits of any of these setups in the current system, however, requires a detailed study and is beyond the scope of this work. We shall only note that an important condition for such interference experiments is the quantum coherence of the fractional bound states at least over the distance they move and the period of observation. While we do not see a general reason forbidding this coherence, attaining it in a real system will present a significant experimental challenge.

\section{Conclusion}\label{sec:conc}

We showed in detail that the proximity of two well-understood, weakly interacting systems can result in some very interesting phenomena. Neither the superconductor nor the integer quantum Hall state in the 2DEG support fractionally charged quasiparticles. Their ground states can be understood by filling a set of single-particle states. However, when brought together, they give rise to excitations that carry fractional charge and obey fractional statistics. This result rests on the following general conditions: (i) the exact quantization of flux by the superconducting film in integer multiples of $\frac12\Phi_0$, half the natural flux quantum of the 2DEG; (ii) the incompressibility of the quantum Hall state of the 2DEG; and (iii) the pinning of the background Abrikosov vortex lattice by the artificial array of pinning sites. Conditions (i) and (ii) are manifestations of gauge invariance and are hence robust against weak perturbation (hence, disorder) in the system. Condition (iii) is necessary to keep the vortex lattice from adjusting itself when a quantized flux moves, basically allowing defects to be created. Otherwise, the incompressible vortex lattice will rearrange and thus screen any excess charge from accumulating in the 2DEG.

This general basis is what connects our different studies in previous sections. The tight-binding model is relevant for strong Zeeman coupling due to large values of the gyromagnetic ratio.~\cite{BerRapJan05a} The continuum model of spin-polarized electrons with uniform background field is relevant for large values of penetration depth, $\ell_B\ll\lambda_L$, whereas the intermediate regime was explored by the continuum model of Pauli electrons with $g=2$. Finally, the effective theory formulation allowed us to properly account for the role of the superconductor and clarify the origin of the length scale $\xi$, and $\xi_0$ in Sec.~\ref{sec:cont}.

A conceptual subtlety must be noted in our usage of the terms ``quasiparticle'' and ``excitation'' for the bound states produced by the vortex lattice defects. These bound states are clearly not excitations of the 2DEG in isolation. Indeed, inserting a flux $\pm\frac12\Phi_0$ changes the Hamiltonian, and hence the energy spectrum, of the 2DEG. A true excitation of the isolated 2DEG is only found by inserting a flux $\pm\Phi_0$, the natural flux quantum for unpaired electrons. However, these bound states are true excitations of the system as a whole in much the same way as the fractionally charged domain walls of the polyacetylene chain are true excitations of the 1D electron-phonon system.~\cite{SuSchHee79a} In both cases, the electronic system in the 2DEG or 1D chain responds to defects in an ordered state arising from external interactions, in our case in the supercondcuting film and in the polyacetylene case in the ionic system. This is also clearly seen in the effective theory formulation of Sec.~\ref{sec:eff}: a vortex defect in the vortex-lattice phase is a finite-energy excitation of the whole system, shown to have fractional charge and statistics.

We briefly comment on the range of parameters necessary to realize our proposed system in the laboratory. Several length scales must be considered: the penetration depth $\lambda_L$, the coherence length $\xi_0$, the magnetic length $\ell_B$, the thickness $d$ of the superconducting film, the separation $w$ between the superconductor and the 2DEG, and the size $L$ of the system. A natural order between some of these scales must exist in order to realize the physics discussed in this paper. Specifically, we need $\xi_0\ll d\lesssim\lambda_L\ll L$. The requirement $\xi_0\ll \lambda_L$ ensures the superconductor is a strong type-II with high enough $H_{c2}\gg B$ the external field needed to realize the quantum Hall state in the 2DEG. In high-temperature cuprate superconductors $\lambda_L\sim$ 1000 \AA\  and $\xi_0\sim$ 10 \AA\ and $H_{c2}$ can go beyond 100 T (Ref.~\onlinecite{Kra98a}). Typical values of $B$ are 5--15 T in semiconductor heterostructures.~\cite{WilEisSto87a} This corresponds to $\ell_B\sim$ 150 \AA, which determines the average spacing of the artificial lattice of pinning sites. This value is not too far from what is achievable in today's artificial structures~\cite{WelXiaNov05a,EisOetPfa07a} and can be increased if lower values of $B$ are possible. Also, one could partially pin only a subset of the Abrikosov lattice using a larger lattice spacing with a specially engineered shape. For a large enough subset and the right shape, the rest of the vortex lattice will be pinned by vortex-vortex interactions. Defects may be created and moved the easiest when $d\to0$. However, the bulk penetration depth is modified in a thin film, given by the effective Pearl penetration depth $\lambda_{\mathrm{eff}}=\lambda_L^2/d$, and diverges as $d\to 0$. This would destroy the bound state. Therefore the optimal conditions to allow the creation and motion of defects as well as the corresponding bound states are found when $d\lesssim\lambda_L$. We also require the size of the system to be larger than all other length scales to enhance the 2D nature of the system and minimize any field leakage from the edges of the system. Ideally, the separation $w\ll \lambda_L$ so that most of the field that exists at the surface of the superconductor enters the 2DEG. However, we expect our conclusions to stand with minor corrections up to $w\lesssim\lambda_L$.

The free electrons with gyromagnetic ratio $g=2$ considered in Sec.~\ref{sec:free} may not be just a theoretically convenient limit. Such electrons are realized on the surface of liquid helium and have been under investigation for some time.~\cite{MonKon04a} The typical electron density and mobility has so far been generally smaller than those in the semiconductor heterostructures and, to the best of our knowledge, a quantum Hall state has not been obtained yet.

Finally, we note that the same set-up can be used with the {\em fractional} quantum Hall states of the 2DEG. For instance, when $\nu=5/2$, by the same general argument leading to Eq.~(\ref{eq:dQgen}) we find
\begin{equation}
\delta Q_{5/2}=\frac e4.
\end{equation}
This is the same charge carried by the true excitations of the Moore-Read state,~\cite{MooRea91a,ReaGre00a} thought to describe the ground state at $\nu=5/2$. This is consistent with the fact that the Moore-Read state can be thought of as a paired-state of electrons (corresponding to a $p+\ii p$ superconductor with a fixed particle number), which implies the natural flux quantum in this state is indeed $\frac12\Phi_0$. We then predict that the bound states found in such a set-up will behave as non-Abelian anyons.~\cite{NayWil96a} If so, our proposals for detecting the fractional charge and the all-electric anyon shuttle in Sec.~\ref{sec:exp} gain new significance, as they can now be used to measure the fractional charge $e/4$ and \emph{braid non-Abelian anyons}.

\section*{Acknowledgment}

The authors acknowledge useful discussions with J. Eckstein, C. W. Hicks, N. Mason, M. M. Vazifeh and S. Vishveshwara.  This work has been supported in part by NSERC, \nobreak{CIfAR}, the Killam Foundation, and the Institute for Condensed Matter Theory at the University of Illinois at Urbana-Champaign.

\appendix            

\section{Schr\"odinger equation with $\delta$-function flux}\label{app:Schr}
The Schr\"odinger equation $\mathcal{H}_{\mathrm{pol}}\Psi=E\Psi$ for Eq.~(\ref{eq:SchrH}) is solved by a wavefunction $\Psi=e^{\ii l\varphi}R(r)$ in the dimensionless polar coordinate introduced in Sec.~\ref{sec:contpol}. In terms of the dimensionless energy $\epsilon = 2E/\hbar\omega$, where $\omega=eB_0/m_ec$ is the cyclotron frequency ($m_e\ell_B^2\omega=\hbar$), we find
\begin{equation}
R''+\frac1r R'-\left[\frac{(l-\eta)^2}{r^2}-r^2+2(\epsilon+l-\eta)\right]R=0.
\end{equation}
Then by the usual Frobenius series expansion we can find the energy levels to be
\begin{equation}
\epsilon_{k,l}^{(\eta)}=2k+|l-\eta|-(l-\eta)+1,
\end{equation}
with the corresponding normalizable eigenstate
\begin{equation}
R_{k,l}^{(\eta)}(r)=r^{|l-\eta|}e^{-\frac{r^2}{2}}\sum_{j=0}^k C_{2j}^{(\eta)}r^{2j},
\end{equation}
and the coefficients 
\begin{equation}
C_{2j}^{(\eta)}=C_0^{(\eta)}\prod_{s=0}^{j-1}\frac{2(s-k)}{(s+1)(2s+2|l-\eta|+3)}.
\end{equation}

So for $l\geq\eta$ we find the levels $E_{k,l}^{(\eta)}=\hbar\omega(k+\frac12)$, which are the usual Landau levels. But for $l<\eta$ we have $E^{(\eta)}_{k,l}=\hbar\omega(k-l+\frac12)+\eta\hbar\omega$, which are located inside the gaps between the Landau levels with a degeneracy $k-l+1$. These states carry an angular momentum opposite to that of the usual Landau levels. This pushes their energy up but in the presence of the fractional extra flux this increase is not a whole integer multiple of the level spacing $\hbar\omega$. We recover the full Landau level structure when $\eta=0$ or $1$.

The normalized states in the LLL are given by
\begin{equation}\label{eq:LLLpsi}
\Psi_l^{(\eta)}(z)=\frac{|z|^{-\eta}z^le^{-\frac12{|z|^2}}}{\sqrt{2\pi \ell_B^2\Gamma(1+l-\eta)}},
\end{equation}
where the complex coordinate $z=re^{\ii\varphi}$.

\section{Aharonov-Casher solution}\label{app:AC}
The Hamiltonian~(\ref{eq:freeH}) can be written 
\begin{equation}
\mathcal{H}_{\mathrm{free}}=\frac1{2m_e}(\sigma_x\Pi_x+\sigma_y\Pi_y)^2,
\end{equation}
where $\boldsymbol{\Pi}=\vex p-\frac ec \vex A$ is the dynamical momentum operator, and we have used the commutation relation $[\Pi_x,\Pi_y]=\ii e\hbar B/c.$ The ground states are zero energy states, $\mathcal{H}_{\mathrm{free}}\psi_0=0$ found by solving $(\sigma_x\Pi_x+\sigma_y\Pi_y)\psi_0=0$. Obviously $\psi_0$ must be an eigenstate of $\sigma_z$. Renormalizable solutions only exist when the spin is aligned with the magnetic field, i.e. $\sigma_z\psi_0=\psi_0$, as physically expected. In the Coulomb gauge  $\nabla\cdot\vex A=0$, we may write the vector potential in terms of a scalar $\phi$, $\vex A=\frac{\Phi_0}{2\pi}(-\partial_y\phi,\partial_x\phi)$, which satisfies
\begin{equation}\label{eq:nabla2phi}
\nabla^2\phi = \frac{2\pi}{\Phi_0}B.
\end{equation}
The zero-mode equation takes the form,
\begin{equation}
\frac{\d\psi_0}{\d \bar z} + \frac{\d\phi}{\d \bar z}\psi_0=0,
\end{equation}
where $z=x+iy$ is the (dimensionful) complex coordinate and $\bar z$ is the complex conjugate. This can be solved by
\begin{equation}
\psi_0=f(z)e^{-\phi},
\end{equation}
where $f$ is an entire function of $z$, $\d f/\d \bar z=0$. 

For a uniform field, we have
\begin{eqnarray}
\phi&=&\frac1{4\ell_B^2} |z|^2 ~~~~~~~~~ \mathrm{(symmetric\ gauge),}\label{eq:phisymm}\\
\phi&=&\frac1{2\ell_B^2} (y^2\ \mathrm{or}\ x^2) ~~~~ \mathrm{(Landau \ gauge),} \label{eq:phiL}
\end{eqnarray}
Choosing $f(z)=z^l$ in the symmetric gauge, we find at once the LLL states ($\eta=0$ in Appendix~\ref{app:Schr}), where $l=0,1,\cdots$ indexes the angular momentum.

\section{TKNN invariant for the Dirac Hamiltonian}\label{app:tknn}
The negative-energy eigenstate of the massive Dirac Hamiltonian~(\ref{tknn2}) can be written as
\begin{equation}
\psi_{\bf p}={1\over \sqrt{2}}\left(
\begin{array}{c}
-\phi_{\bf p}\sqrt{1-m/E_{\bf p}} \\
\sqrt{1+m/E_{\bf p}}
\end{array}
\right),
\end{equation}
where $\phi_{\bf p}=(p_x-\ii p_y)/|\vex p|$. 
Explicit computation using Eq.~(\ref{tknn12}) gives
\begin{equation}
\mathcal{A}_{\bf p}=\frac12\frac{\hat{z}\times \vex p}{E_{\vex p}(m+E_{\vex p})}.
\end{equation}
It is easiest to evaluate Eq.\ (\ref{tknn1}) using the Stokes theorem,
\begin{eqnarray}\label{tknn7}
\mathcal{K}^\pm &=& \frac1{2\pi}\int_{\mathrm{BZ}}\d^2\vex p\:(\nabla_{\bf p}\times {\cal A}_{\bf p})\cdot\hat{z}\\ \nonumber
 &=& \frac1{2\pi}\int_{\mathrm{BZ}}\d^2\vex p\: \frac{m}{2E_{\vex p}^{3}}= \frac12\mathrm{sgn}(m),
\end{eqnarray}
where, 
 in the last step, we have extended the upper bound of the integral to infinity. The latter approximation is accurate as long as $|m|\ll W$ (the bandwidth), and becomes exact in the limit $m\to 0$
when the Berry flux becomes a delta function at the node. Since $\mathcal{K}$ is a topological invariant, constrained to be an integer, it cannot change as $m$ is varied as long as the gap remains open. The above calculation thus gives the exact result.


\begin{thebibliography}{10}

\bibitem{Wil82a}
F.~Wilczek,
\newblock \prl {\bf 49}, 957 (1982).

\bibitem{LeiMyr77a}
J.~M. Leinaas and J.~Myrheim,
\newblock Il Nuovo Cimento {\bf 37B}, 1 (1977).

\bibitem{JacReb76a}
R.~Jackiw and C.~Rebbi,
\newblock \prd {\bf 13}, 3398 (1976).

\bibitem{SuSchHee79a}
W.~P. Su, J.~R. Schrieffer, and A.~J. Heeger,
\newblock \prl {\bf 42}, 1698 (1979).

\bibitem{GolWil81a}
J.~Goldstone and F.~Wilczek,
\newblock \prl {\bf 47}, 986 (1981).

\bibitem{Kit03a}
A.~Y. Kitaev,
\newblock Ann. Phys. {\bf 303}, 2 (2003).

\bibitem{freedman1}
M.~Freedman, C. Nayak, K. Shtengel, K. Walker and Z. Wang,
\newblock Ann. Phys. {\bf 310}, 428 (2004).

\bibitem{Lau83b}
R.~B. Laughlin,
\newblock \prl {\bf 50}, 1395 (1983).

\bibitem{AroSchWil84a}
D.~Arovas, J.~R. Schrieffer, and F.~Wilczek,
\newblock \prl {\bf 53}, 722 (1984).

\bibitem{Wil82b}
F.~Wilczek,
\newblock \prl {\bf 48}, 1144 (1982).

\bibitem{HouChaMud07a}
C.-Y. Hou, C.~Chamon, and C.~Mudry,
\newblock \prl {\bf 98}, 186809 (2007).

\bibitem{SerWeeFra08a}
B.~Seradjeh, C.~Weeks, and M.~Franz,
\newblock \prb {\bf 77}, 033104 (2008).

\bibitem{ChaHouJac07a}
C.~Chamon {\it et al.},
\newblock \prl {\bf 100}, 110405 (2008).

\bibitem{SerFra08a}
B.~Seradjeh and M.~Franz,
\newblock \prl {\bf 101}, 146401 (2008).

\bibitem{WeeRosSer07a}
C.~Weeks, G.~Rosenberg, B.~Seradjeh, and M.~Franz,
\newblock Nature Phys. {\bf 3}, 796 (2007).

\bibitem{KirTsuRup96a}
J.~R. Kirtley, C.~C. Tsuei, M.~Rupp, J.~Z. Sun, L.~S. Yu-Jahnes, A.~Gupta, M.~B. Ketchen, K.~A. Moler, and M.~Bhushan,
\newblock \prl {\bf 76}, 1336 (1996).

\bibitem{AusLuaStr08a}
O.~M. Auslaender, L.~Luan, E.~W.~J. Straver, J.~E. Hoffman, N.~C. Koshnick, E.~Zeldov, D.~A. Bonn, R.~Liang, W.~N. Hardy, and K.~A. Moler,
\newblock Nature Phys. {\bf 5}, 35 (2009).

\bibitem{BluSebGui06a}
H.~Bluhm, S.~E. Sebastian, J.~W. Guikema, I.~R. Fisher, and K.~A. Moler,
\newblock \prb {\bf 73}, 014514 (2006).

\bibitem{GarWynBon02a}
B.~W. Gardner, J.~C. Wynn, D.~A. Bonn, R.~Liang, W.~N. Hardy, J.~R. Kirtley, V.~G. Kogan, and K.~A. Moler,
\newblock Appl. Phys. Lett. {\bf 80}, 1010 (2002).

\bibitem{StrHofAus08a}
E.~W.~J. Straver, J.~E. Hoffman, O.~M. Auslaender, D.~Rugar, and K.~A. Moler,
\newblock Appl. Phys. Lett. {\bf 93}, 172514 (2008).

\bibitem{ConSch74a}
E.~Conen and A.~Schmid,
\newblock J. Low Temp. Phys.  {\bf 17}, 331 (1974).

\bibitem{WelXiaNov05a}See e.g.\
U.~Welp, Z.~L. Xiao, V.~Novosad, and V.~K. Vlasko-Vlasov,
\newblock \prb {\bf 71}, 014505 (2005) and references therein.

\bibitem{wilczekbook} F. Wilczek, Fractional statistics and anyon 
superconductivity (World Scientific, Singapore, 1990). 

\bibitem{AhaCas79a}
Y.~Aharonov and A.~Casher,
\newblock \pra {\bf 19}, 2461 (1979).

\bibitem{Brandt} 
E.H. Brandt, 
\newblock J. Low.\ Temp.\ Phys.\ {\bf 24}, 709 (1977) and {\bf 73}, 355, (1988);  \prb {\bf 37}, 2349 (1988).

\bibitem{DuhVet90a}
P.~Duhamel and M.~Vetterli,
\newblock Signal Processing {\bf 19}, 259 (1990).

\bibitem{BerRapJan05a}
M.~Berciu, T.~G. Rappoport and B.~Janko,
\newblock Nature {\bf 435}, 71 (2005).

\bibitem{tknn}
D.J. Thouless, M. Kohmoto, M.P. Nightingale, and M. den Nijs, 
\newblock \prl {\bf 49}, 405 (1982).

\bibitem{SuSchHee80a}
W.~P. Su, J.~R. Schrieffer, and A.~J. Heeger,
\newblock \prb {\bf 22}, 2099 (1980).

\bibitem{Bar64a}
V.~Bargmann,
\newblock J. Math. Phys. {\bf 5}, 862 (1964).

\bibitem{SimMuk93a}
R.~Simon and N.~Mukunda, 
\newblock \prl {\bf 70}, 880 (1993).

\bibitem{RabArvMuk99a}
E.~M. Rabei, Arvind, N.~Mukunda, and R.~Simon,
\newblock \pra {\bf 60}, 3397 (1999). 

\bibitem{Ger89a}
Ph.~de~Sousa Gerbert,
\newblock \prd {\bf 40}, 1346 (1989).

\bibitem{Wen04a}
X.-G. Wen,
\newblock {\em Quantum Field Theory of Many-body Systems} (Oxford University Press, 2004).

\bibitem{Pes78a}
M.~E. Peskin,
\newblock Ann. Phys. {\bf 113}, 122 (1978).

\bibitem{NagLee00a}
N.~Nagaosa and P.~A. Lee,
\newblock \prb {\bf 61}, 9166 (2000).


\bibitem{goldman1} V.J. Goldman and B. Su, Science {\bf 267}, 1010 (1995).

\bibitem{saminadayar1} L. Saminadayar, D.C. Glattli, Y. Jin, and B. Etienne, \prl {\bf 79}, 2526 (1997).

\bibitem{picciotto1} R. De-Picciotto {\em et al.}, Nature {\bf 389}, 162 (1997).

\bibitem{heeger1} A.J. Heeger, S. Kivelson, J.R. Schrieffer, and W.-P. Su, 
Rev. Mod. Phys. {\bf 60}, 781 (1988).

\bibitem{camino1} F.E. Camino, Wei Zhou, and V.J. Goldman, \prb {\bf 72}, 075342 (2005).

\bibitem{sarma1} S. Das Sarma, M. Freedman, and C.Nayak, \prl {\bf 94}, 166802 (2005).

\bibitem{bonderson1} P. Bonderson, A. Kitaev, and K. Shtengel, \prl {\bf 96}, 016803 (2006).

\bibitem{Kra98a}
H. Krauth,
\newblock in {\em Handbook of Applied Superconductivity}, vol. 1, ed. B. Seeber (IOP, London, 1998).

\bibitem{WilEisSto87a}
R. Willett {\it et al.},
\newblock \prl {\bf 59}, 1776 (1987).

\bibitem{EisOetPfa07a}
J.~Eisenmenger,  M.~Oettinger, C.~Pfahler, A.~Plettl, P.~Walther, and P.~Ziemann,
\newblock \prb {\bf 75}, 144514 (2007).

\bibitem{MonKon04a}
For a review see Yu. P. Monarkha and K. Kono,
\newblock {\em Two-Dimensional Coulomb Liquids and Solids} (Springer, Berlin, 2004).

\bibitem{MooRea91a}
G.~Moore and N.~Read,
\newblock Nucl. Phys. B {\bf 360}, 361 (1991).

\bibitem{ReaGre00a}
N.~Read and D.~Green,
\newblock \prb {\bf 61}, 10267 (2000).

\bibitem{NayWil96a}
C.~Nayak and F.~Wilczek,
\newblock Nucl. Phys. B {\bf 479}, 529 (1996).


\end{thebibliography}

\end{document}